\documentclass{aa}  

\usepackage{graphicx}
\usepackage{caption}
\usepackage{hyperref}

\begin{document}

   \title{Preferential accretion of binary stars}

   \subtitle{}

   \author{D. Dalsgaard
          \inst{1}
          \and
          M. Kuffmeier
          \inst{1}
          \and
          T. Haugb{\o}lle \inst{1}
          }

   \institute{Niels Bohr Institute, University of Copenhagen, Jagtvej 155 A, DK-2200 Copenhagen, Denmark,
   \email{xdz546@alumni.ku.dk, kueffmeier@nbi.ku.dk, haugboel@nbi.ku.dk}
   }

   \date{Received \today}

\abstract{
The attracting properties of gravity enable matter present in dense cores to collapse into stars with seven orders of magnitude change in space and time making modelling of star formation a challenging multi-scale process. To circumvent this scale problem stars are replaced by a sub-grid sink particle at a much larger scale. Sink particles are created above a threshold density and acquire mass and momentum through accretion. In models where binary star systems form and migrate to separations of a few cells, the accretion flow is unresolved and the relative accretion rate to the sink particles may become inaccurate. We introduce a new recipe for accretion onto binary sink particles that have overlapping accretion regions and implement an algorithm to track the angular momentum of sink particles as a proxy for the stellar spin. Our preferential binary accretion recipe uses a virtual binary sink particle for the purpose of accretion and redistribute the accreted mass onto the sink particles according to results from models investigating binary accretion in detail. This solves problems common to current algorithms in many codes: (i) accretion is not suppressed due to large velocity differences between gas and stars, when that velocity is only internal to the binary system, (ii) the accretion rates are smoother for the unresolved close binaries in eccentric orbits, and (iii) non-physical suppression of accretion onto the secondary sink particle when the primary dominates the potential is eliminated. We test our implementation by comparing simulations at increasing resolution until the binaries are resolved. While not perfect, the algorithm mitigates undesired properties of current algorithms and is particularly useful for global models of star-forming regions. It may also be applied to other unresolved accreting binaries, such as compact objects in evolved star clusters and binary supermassive black holes in cosmological models.}


\maketitle

\section{Introduction}
The always-attracting properties of self-gravity make matter from dense cores up to around 10 000 AU collapse into stars that are only 0.001 AU in size.
The cores are not isolated, and during the formation process turbulent and filamentary streamers originating at up to millions of AU away fuel the process \citep[][]{Pineda+2023}.
This makes star formation a multi-scale and multi-physics problem requiring modelling over more than 7 orders of magnitude in space and time.
Even though recent codes with sophisticated algorithms, like adaptive mesh refinement, can span such a dynamic range \citep{Kuffmeier_episodic_2018} and follow the collapse of matter in a single, isolated core to stellar densities for short time spans of up to a thousand years \citep[e.g.,][]{Wurster+2020}, it is still not possible to do so for full star-forming regions.
To limit the cost of simulating the dynamics to ever-smaller time and spatial scales, over the past three decades a sub-scale model for stars has been developed.
Sink particles are introduced where gas collapse to densities so high that fragmentation begins to be dominated by grid-scale effects.
Similar but slightly different approaches have been used to determine the exact criteria for the formation and the accretion of gas to the sink particle \citep[see e.g.][]{Bate1995,Belyakov2004,Federrath2010,Springel2010,Gong2013,Bleuler2014,Hopkins2015,Haugbolle2018,Vorobyov+2021}.
What all of these have in common is that they numerically integrate self-gravitating hydrodynamics together with additional physics, such as magnetic fields, heating and cooling, radiative transfer, and a sub-grid model for stellar particles.
They have been used with success to study the evolution, statistics, star formation efficiency, and multiplicity properties of star-forming regions.

In principle, the accuracy of a sink particle algorithm is easy to ascertain: at resolved scales it should lead to the same results as if no sink particles are employed and the gravitational collapse followed ab initio to stellar length scales and densities. In practice, this is highly non-trivial for two reasons: (i) protostars often form in multiple or binary systems \citep{Duquennoy1991,Connelley2008,Raghaven2010,Offner2022}. If the distance at which such stars form is not resolved in the simulations, the full multiplicity statistics will not be recovered. (ii) young stars are universally observed to drive outflows, and the outflow speed and mass ejection rate depend on the rotational energies available in the disc. Comparing models with different smallest cell size will therefore resolve the outflows to a different degree and result in different stellar masses and kinematic feedback. The combined resolution dependence of multiplicity and outflows makes it very difficult to study the convergence properties of global models of star-forming regions that include sink particles and do not reach a resolution of less than $\approx 10^{-2}$ AU. Sink particle recipes contain two parts: (i) criteria for when to form a new sink particle and optionally merge sink particles, and (ii) a model for how gas is accreted to the sink particle. In addition, they interact gravitationally with the gas, and may include different algorithms for stellar feedback such as radiation, outflows, stellar winds and supernova explosions. The properties of the stellar feedback depends sensitively on the binarity and mass evolution of the sink particles.

We study the accretion properties in a model of a star-forming region carried out with the adaptive mesh refinement code {\sc ramses} at different resolutions and introduce a new recipe to better capture the accretion flow of gas to close binaries with overlapping accretion regions. The recipe is based on theoretical models from detailed studies of accretion from a circumbinary disc to binary black holes by \cite{Kelley2019} and \cite{Siwek2023}. In section 2 we introduce the methods, in section 3 the model, and in section 4 the results of our simulations. The results are put in perspective in section 5 and we end with the conclusions.

\section{Methods} \label{methods}

\subsection{Magnetohydrodynamics}\label{subsection: MHD}

The interstellar medium (ISM) carries significant weight, is magnetised, and is charged. A minimal description therefore must include self-gravitating magnetohydrodynamics (MHD) as outlined in Eqs.~\ref{eq: MHD}. These are written in conservative form as implemented in the finite volume formulation of {\sc ramses}~\citep{Fromang2006}. Additionally, the Poisson equation is solved to compute the gravitational force-density as described below in section \ref{subsection: modelling gravity}.

\begin{equation}\label{eq: MHD}
\begin{aligned}
& \frac{\partial \rho}{\partial t}+\mathbf{\nabla} \cdot(\rho \mathbf{v}) =0 \\
& \frac{\partial \rho \mathbf{v}}{\partial t}+\mathbf{\nabla} \cdot(\rho \mathbf{v} \otimes \mathbf{v}-\mathbf{B}  \otimes \mathbf{B})+\mathbf{\nabla} P_{\textrm{tot}} = -\rho \nabla \Phi \\
& \frac{\partial E}{\partial t}+\mathbf{\nabla} \cdot\left[\left(E+P_{\textrm{tot}}\right) \mathbf{v}-\mathbf{B}(\mathbf{B} \cdot \mathbf{v})\right] = -\rho \mathbf{v}\nabla \Phi \\
& \frac{\partial \mathbf{B}}{\partial t} = \mathbf{\nabla} \times (\mathbf{v} \times \mathbf{B}) \\
& \nabla^2 \Phi=  4 \pi\, \textrm{G}\, \rho\,,
\end{aligned}
\end{equation}
where $\mathbf{v}$ is the velocity of the fluid, $\rho$ is the density of the fluid, $\mathbf{B}$ is the magnetic field, and $P_\textrm{tot}$ includes the thermal and magnetic pressure
\begin{equation} \label{eq: total pressure}
    P_\textrm{tot} = P + \frac{\mathbf{B}\cdot \mathbf{B}}{2}\,,
\end{equation}
$E$ is the total energy density of the fluid
\begin{align} \label{eq: total energy}
    E & = E_\textrm{therm} + E_\textrm{kin} + E_\textrm{mag} \\
    & = \rho \epsilon + \rho \frac{\mathbf{v \cdot v}}{2} + \frac{\mathbf{B}\cdot \mathbf{B}}{2}\,,
\end{align}
where $\epsilon$ is the specific internal energy. $\Phi$ is the gravitational potential as determined by the Poisson equation for a given mass density.

The thermal pressure is related to the internal energy and density by an adiabatic equation of state $P = (\gamma - 1) \rho \epsilon$, where $\gamma$ is the adiabatic index. In our study we choose an adiabatic index of $\gamma=1$, making the gas isothermal. The solenoidal condition, $\nabla \cdot \mathbf{B} = 0$, is maintained to round-off precision in all cells by defining the magnetic field at cell interfaces and updating it using constrained transport \citep{Fromang2006}.

\subsection{Self-gravity} \label{subsection: modelling gravity}
The gravitational acceleration is computed with a split formalism as introduced by \citet{Haugbolle2018} with three different contributions. The combined density of the gas and sink particles is used to solve the Poisson equation (Eq.~\ref{eq: MHD}) and obtain $\Phi_{\textrm{gas+sink}}$ and the related gravitational acceleration on the gas. The density of the gas only is used as input for the Poisson equation to obtain $\Phi_{\textrm{gas}}$, which then allows the calculation of the gravitational acceleration due to the gas on the sink particles. Finally, the direct Newtonian acceleration of the sink particles, modulated with a cubic spline softening at the cell scale, is used to obtain the pair-wise accelerations of the sink particles.

The two main limitations are that for binaries' separated by less than a tenth of a cell no further migration to smaller distances happens, due to the softening of the pair-wise acceleration, and that the acceleration induced by sink particles is computed slightly different for the gas and the sinks, which may induce numerical fictitious relative accelerations. The error is usually much lower than the typical differential accelerations induced by magnetic forces and pressure gradients. The advantage of the method is that it scales to tens of thousands of sink particles while being capable of evolving multiple star systems, and the smoothing makes the time-step criterion from the sink particle orbital speed similar to the MHD time step.

\subsection{Sink particles} \label{subsection: sink particles}
Simulating star formation pushes the boundaries of available computational resources and wall-clock integration time, because of the vast scale it encompasses, from the ISM to the interior of stars themselves. Modelling becomes increasingly challenging as the density increases in the protostellar collapse, dominating computational time and reducing the time step to such an extent that continuing the simulation becomes impractical shortly after the formation of the star. To get around these limitations, we adopt a strategy in which, upon reaching a density threshold at the smallest cell size corresponding to resolving the Jeans length by a few cells, a sink particle is inserted. This concept, first introduced to simulate star formation by \citet{Bate1995}, significantly reduces computational demands. It assumes that we can make a cut-off in the model and encapsulate the internal dynamics and evolution of the protostar as a collision-less particle that interacts with the fluid according to an ad hoc algorithm. It enables us to continue the simulation of the broader system on dynamically interesting timescales. The specific sink algorithm we use is described in-depth in \citet{Haugbolle2018} and is outlined below.
In addition, we introduced the tracking of sink particle spin and consideration of binary properties when accreting onto sink particles with overlapping accretion regions.
Previous estimates of the stellar spin in simulations using the code were obtained by post-processing of tracer particles prior to accretion onto the sink \citep{Kuffmeier+2024,Pelkonen+2025}, which required follow-up analysis. 
To track the spin of the sinks, we implemented an optimized version of the accretion recipe by \citep{Federrath2014}.

\subsubsection{Sink creation}
A sink particle is inserted in cells where all the following conditions are true
\begin{enumerate}
    \item The density of the gas exceeds a predefined threshold, denoted $\rm \rho_s$. Tests in \cite{Haugbolle2018} have shown that a threshold density equivalent to resolving the Jeans length with two cells is optimal for suppressing formation of spurious sink particles at non-collapsing transient density peaks, while simultaneously allowing for formation of sink particles in bona fide collapsing density peaks.
    \item The gravitational potential must be a local minimum, when computed at the second highest level of refinement and compared to the 26 neighbouring cells.
    \item The velocity is converging, $\nabla \cdot \mathbf{v} < 0$.
    \item The minimum distance to an existing sink particle is above $r_\textrm{ex}$, typically set to 8 cells.
\end{enumerate}
Sink particles are instantiated with no mass and zero momentum, but immediately acquire mass through accretion.

\subsubsection{Accretion of mass and momentum} \label{subsection: mass and momentum}
Sink particles can accrete mass from surrounding cells within their accretion radius, $\rm r_{acc}$, typically set to half the exclusion radius, $\rm r_{ex}$, for sink creation or 4 cells at the highest level of refinement. Any model for accretion of gas to sink particles is ad hoc, as it happens at a distance, while the physical process of accretion happens at the stellar surface. Furthermore, depending on the resolution of the model, outflows close to the sink particle may or may not be accounted for by the model. In our model, accretion is only allowed from the gas bound to the sink, determined by the relative velocity between the gas and the sink particle compared to the Keplerian velocity $v_K = \left(G m_\textrm{sink}/d\right)^{1/2}$, where $d$ is the distance between the sink particle and the cell. The accretion rate, $\dot{m}_{\textrm{gas}}$, is given by
\begin{equation}\label{eq: accretion rate sinks}
\dot{m} = \frac{\Delta m}{\Delta t}= \begin{cases}\alpha_\textrm{rate } m_\textrm{cell} v_K \Delta x^{-1} f_v & \text { if } \rho \leqslant \rho_\textrm{max,acc} \\ 0.5\,m_\textrm{cell} \Delta t^{-1} & \text { if } \rho>\rho_\textrm{max,acc}\end{cases}\,.
\end{equation}
The accreted mass is related to the change in density of the cell as $\Delta m = \Delta\rho\, \Delta V$. where $\Delta V$ is the cell volume. The second case, when the gas density is above $\rho_\textrm{max,acc}$ -- normally set to twice the threshold density for sink creation -- is a safety valve to disallow the density to reach too high values inside the accretion radius. The accretion efficiency is modulated by $\alpha_{\textrm{rate}}$, which is related to the fraction accreted per orbital time-scale, and by the function $\rm f_v$ (see Eq.~\ref{eq: f_v}), which adjusts accretion based on the ratio of total energy to potential energy and the distance from the sink.
\begin{equation} \label{eq: f_v}
f_v=\left[1-\left(\frac{d}{r_\textrm{acc}}\right)^2\right] \times \begin{cases}0 & \text { if } v \geqslant \sqrt{2} v_K \\ \rm 2-\left(\frac{v}{v_K}\right)^2 & \text { if } v_K<v<\sqrt{2} v_K \\ 1 & \text { if } v \leqslant v_K\end{cases}
\end{equation}
When the amount of mass to be accreted, $\Delta m$, from a single cell has been determined, the corresponding amount of momentum is deposited onto the sink particle, and the thermal energy in the cell is reduced in proportion to the accreted mass fraction as detailed in Eq.~\ref{eq: sink update} for the sink and Eq.~\ref{eq: cell update} for the cell. The updated quantities are denoted with a prime.
\begin{equation} \label{eq: sink update}
\begin{array}{rrl}
\text { mass: } & m_\textrm{sink }^{\prime} & = m_\textrm{sink }+\Delta m \\
\text { c.o.m.: } & m_\textrm{sink }^{\prime} \mathbf{r}_\textrm{sink }^{\prime} & =m_\textrm{sink } \mathbf{r}_\textrm{sink }+\Delta m\, \mathbf{r}_{\mathrm{cell}} \\
\text { momentum: } & m_\textrm{sink }^{\prime} \mathbf{v}_\textrm{sink }^{\prime} & =m_\textrm{sink} \mathbf{v}_\textrm{sink }+\Delta m \,\mathbf{v}_\textrm{cell}
\end{array}
\end{equation}
where c.o.m.~is the updated centre of mass of the sink particle; i.e.~the change in the sink position due to the addition of the cell mass. Correspondingly, the cell quantities are adjusted as
\begin{equation} \label{eq: cell update}
\begin{array}{rrl}
\text { density: } & \rho_\textrm{cell }^{\prime} & =\rho_\textrm{cell }-\Delta \rho \\
\text { momentum: } & \rho_\textrm{cell }^{\prime} \mathbf{v}_\textrm{cell}^{\prime} & = \rho_\textrm{cell } \mathbf{v}_\textrm{cell}-\Delta \rho\, \mathbf{v}_\textrm{cell } \\
\text { energy: } & E^{\prime} & = E_\textrm{therm}^{\prime} + E_\textrm{kin}^{\prime} + E_\textrm{mag} \\
& & =\frac{\rho - \Delta \rho}{\rho} \left[E_\textrm{therm} + E_\textrm{kin}\right] + E_\textrm{mag}
\end{array}
\end{equation}
Magnetic fields are not accreted, and as such the magnetic energy density $E_\textrm{mag}$ is unchanged when updating $E$. How magnetic fields behave during accretion is not well understood. It is likely that a combination of reconnection events and ohmic dissipation of the magnetic energy happens in nature, leading to the magnetic flux being dragged into the forming star and flux being expelled by the outflow. The solenoidal constraint further complicates any numerical sub-grid model making it non-trivial to decrease the magnetic flux, e.g., in proportion to the mass being accreted.

\subsection{Sink particle spin} \label{subsection: spin methods}
The total amount of angular momentum extracted from the gas is deposited onto the sink particle as a spin as part of the accretion process. The spin is a useful indicator of the stellar rotation axis. Numerically, accretion happens at a distance, and the magnitude of the spin will scale with the Keplerian angular momentum at the accretion radius and therefore be strongly dependent on the resolution. Only if interactions at the stellar surface are accounted for, that may result in exchange of angular momentum through e.g.~winds and magnetic field locking, the spin could be a useful proxy for the stellar rotation rate.

Our implementation of spin closely follows \citet{Federrath2014}. While mathematically identical to the equations used in \citet{Federrath2014}, we have simplified the equations by rewriting them in the reference frame of the sink. The sink particle spin, $\mathbf{S}$, is updated according to
\begin{equation} \label{eq: spin new}
    \mathbf{S}_\textrm{sink}^{\prime} = \mathbf{S}_{\textrm{sink}} + \frac{m_\textrm{sink }\Delta m}{m_\textrm{sink}+\Delta m} (\mathbf{r}_\textrm{sink} - \mathbf{r}_\textrm{cell })\times (\mathbf{v}_\textrm{sink} - \mathbf{v}_\textrm{cell })\,
\end{equation}
where the distances are rewritten as relative distances to support periodic boundary conditions \citep[see also][]{Bleuler2014}. In this paper, we use the spin vector to define a natural frame of reference when visualising the gas in the vicinity of sink particles.

\begin{figure}[htb!]
    \centering
    \includegraphics[width=\columnwidth]{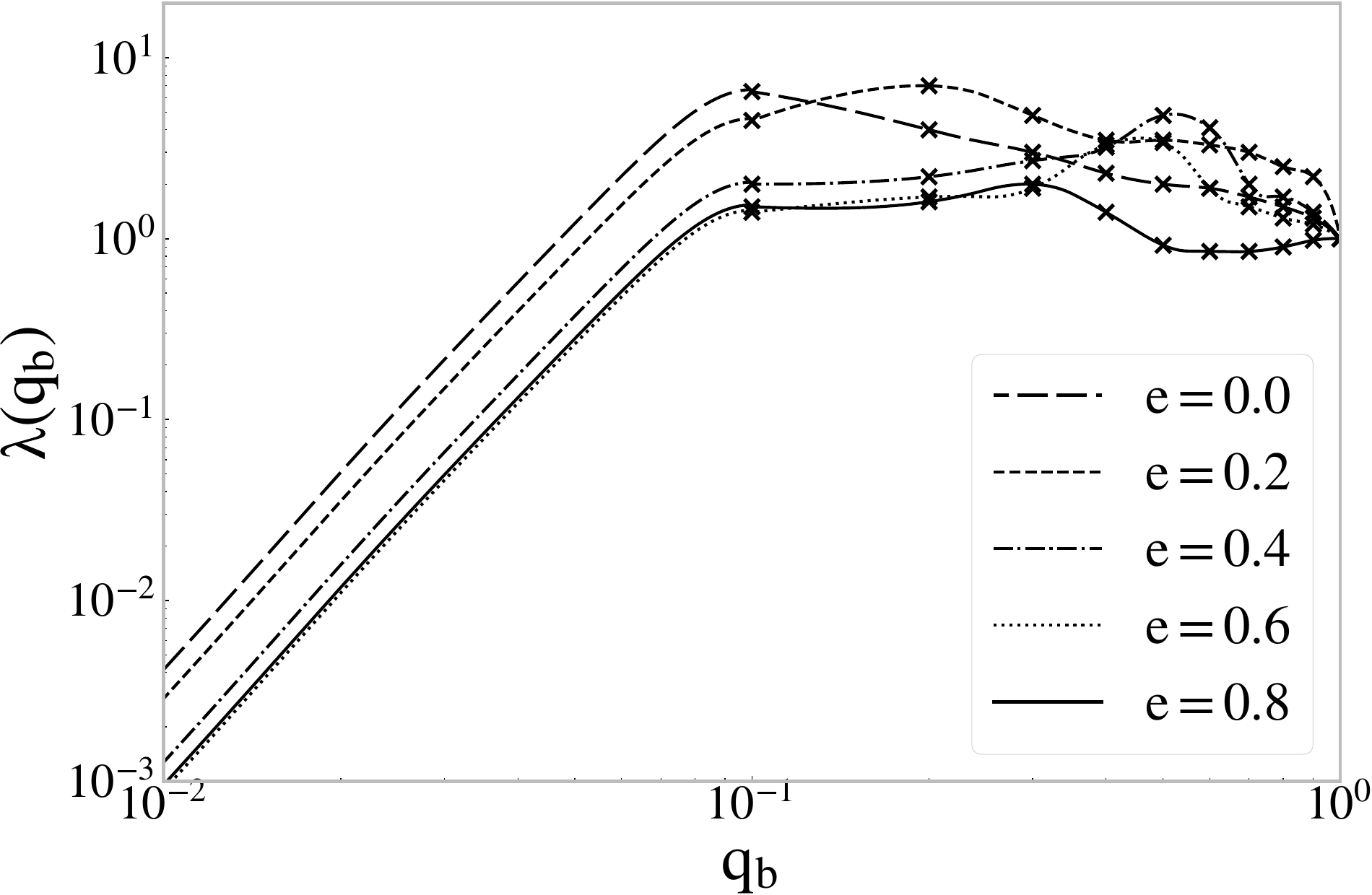}
    \caption{Plot of $\lambda = \dot{m}_s / \dot{m}_p$, the ratio of accretion onto the secondary and primary star, against the mass ratio of in the binary $q_{\rm b}=m_{\rm s} / m_{\rm p}$. The data is merged from \citep{Siwek2023} to the right of $\rm q_{\rm b} = 0.1$ and Eq.~\ref{eq: preferential accretion} which is adapted from \citep{Kelley2019}. The left side of the plot continues to $\rm q_{\rm b} = 0$, but has been limited for a clearer view.}
    \label{fig: preferential accretion}
\end{figure}

\begin{table*}[t]
\centering
\caption{Binary pair information}
\small
\begin{tabular}{|l|l|l|l|l|l|l|l|l|}
\hline
Pair & 75, 82 & 141, 146 & 142, 144 & 217, 219 & 253, 254 & 257, 260 & 279, 281 & 290, 292 \\
\hline
Period [yr] & 149.7 & 139.1 & 1385.8 & 499.0 & 341.8 & 644.4 & 99.1 & 313.8 \\
\hline
Eccentricity & 0.7 & 0.5 & 0.2 & 0.4 & 0.4 & 0.5 & 0.3 & 0.1 \\
\hline
Orbital Energy [erg] & -9.1e+11 & -1.8e+11 & -1.7e+10 & -4.3e+10 & -1.96e+11 & -4.4e+10 & -1.79e+11 & -1.13e+11 \\
\hline
Total Mass [M\_sun] & 14.0 & 1.1 & 0.3 & 0.5 & 3.2 & 0.6 & 0.8 & 1.3 \\
\hline
Mass Ratio & 0.7 & 0.4 & 0.9 & 0.9 & 0.9 & 0.5 & 0.3 & 0.3 \\
\hline
Min Periastron [cell] & 0.5 & 0.3 & 1.4 & 0.6 & 0.8 & 0.7 & 0.3 & 0.8 \\
\hline
Max Apastron [cell] & 2.3 & 0.8 & 1.9 & 1.4 & 2.1 & 1.9 & 0.5 & 1.2 \\
\hline
Distance at levelmax=14 [cell] & 1.6 & 0.6 & 1.7 & 1.0 & 1.6 & 1.4 & 0.4 & 1.0 \\
\hline
Distance at levelmax=15 [cell] & 3.3 & 1.3 & 3.4 & 2.1 & 3.1 & 2.8 & 0.8 & 2.0 \\
\hline
Distance at levelmax=16 [cell] & 6.5 & 2.5 & 6.8 & 4.2 & 6.2 & 5.7 & 1.6 & 4.0 \\
\hline
Distance at levelmax=17 [cell] & 13.0 & 5.0 & 13.6 & 8.4 & 12.4 & 11.4 & 3.2 & 8.0 \\
\hline
Distance at levelmax=18 [cell] & 26.0 & 10.0 & 27.3 & 16.8 & 24.8 & 22.7 & 6.5 & 15.9 \\
\hline
\end{tabular}
\label{table:binary table}
\end{table*}

\subsection{Preferential binary accretion} \label{subsection: preferential accretion}
Accretion of mass and momentum from the gas to sink particles is a sub-grid recipe that has been implemented with some variations in astrophysics by different research groups with the overall goal of reproducing the actual flow of material to the much smaller collapsed object. In general, a limited number of cells inside a given distance, the accretion radius, is considered, and material is removed from the gas and put onto the sink particle according to a prescribed set of rules. Sink particles move through the simulation volume subject to gravitational forces, and may form binary systems. In that case, their accretion regions can overlap for extended periods of time. In our implementation in {\sc ramses}, accretion of mass from a cell is to the closest sink particle \citep{Haugbolle2018}, similar to what is done by \citet{Hubber2013}. In \citet{Federrath2010} and \citet{Bate1995} accretion onto the sink particle happens with the lowest energy, in \citet{Bleuler2014} accretion from a cell is onto the most massive sink particle, in \citet{Grudic2021} entire gas cells are accreted onto the sink with the lowest dynamical time-scale, while other codes merge sink particles when they have overlapping accretion regions \citep{Gong2013,Krumholz2004}.

All these procedures do not reflect the real accretion flow in binary and multiple systems, when the accretion discs remain unresolved, which is a common case in global models of star-forming regions \citep[see e.g.][]{Haugbolle2018}. In particular, if the stars have highly eccentric orbits and large orbital velocities accretion may be suppressed and individual gas cells may deposit mass periodically to the different sinks resulting in a rapidly varying accretion rate \citep{Padoan2014,Jensen2018}. Both observationally and theoretically, it is well-known that the secondary may accrete more efficiently than the primary leading to equal-mass binaries \citep[e.g.][]{LaiMunoz2023}, a process that is at odds with the common choice implemented in many codes, as described above, of accretion onto the most massive or bound sink particle.

To achieve a more realistic accretion rate for unresolved multiple systems we propose a new recipe:
\begin{itemize}
\item{For each cell we check if it is within the accretion radius of multiple sink particles.}
\item{The two sink particles closest to the cell are used, and binary system parameters are computed, including total mass, centre of mass, centre of mass velocity, and eccentricity and mass ratio.}
\item{Mass accretion then proceeds according to the single star recipe, but using the system parameters as input, and the accreted mass, momentum, and spin is distributed to the two sink particles in a fractional manner according to a sub-grid recipe.}
\end{itemize}
This approach eliminates the orbital peculiar velocities allowing for a more accurate assessment of whether the gas is gravitationally bound to the system, resulting in smoother accretion profiles, and it can account for preferential accretion onto the secondary in the sub-scale model for binary accretion. A limitation is that it does not apply to un-resolved triple or higher multiplicity systems. While it would be possible and relatively straight forward to make a hierarchical algorithm, we find that unresolved higher multiplicity systems tend to be both very rare and short-lived.

The sub-grid recipe use literature research on more general preferential binary accretion. Specifically by \citet{Siwek2023}, who has done simulations of binary black holes with mass ratios of $q_{\rm b}=m_{\rm s} / m_{\rm p} = 0.1$ to $q_{\rm b}=1$ for a range of eccentricities reproduced in Fig.~\ref{fig: preferential accretion}, and by \citet{Kelley2019}, who describe preferential accretion for massive black hole binaries at the centre of active galactic nuclei in their Eq.\ 1; an equation which is a fit to simulation results by \citet{Farris2014}. To cover all mass ratios, from $0$ to $1$, we combine Eq.\ 1 from \citet{Kelley2019} with the data from the black hole simulations in \citet{Siwek2023} using a mass ratio of $q_{\rm b} = 0.1$ as the transition point. The equation presented in \citet{Kelley2019} is more complicated, but in the domain $q_{\rm b} \leq 0.1$ it can be simplified to
\begin{equation}\label{eq: preferential accretion}
    \lambda(q_{\rm b}) = \frac{a_3}{(a_4 q_{\rm b})^{a_5} + (a_4 q_{\rm b})^{-a_5}}\,,
\end{equation}
where $\lambda$ is defined as the ratio of secondary and primary accretion rates
\begin{equation} \label{eq: lambda def}
    \lambda(q_{\rm b}) = \frac{\dot{m}_{\rm s}}{\dot{m}_{\rm p}}
\end{equation}
The numerical coefficients are $a_3=50, a_4=10, a_5=3.5$. Notice that we changed $a_4$ slightly from 12, used by \citet{Kelley2019}, to 10 in order to have a smooth transition to the data from \cite{Siwek2023}.

Using these data, we create a table with values corresponding to the curves in Fig.~\ref{fig: preferential accretion} in the range $0 \leq q_{\rm b} \leq 1$ and with eccentricities in the range $0 \leq e \leq 0.8$. For systems with larger eccentricities we use the value for $e=0.8$ in our table.

The algorithm has been implemented into our version of {\sc ramses} and is available online\footnote{\url{https://erda.dk/cert_redirect/public_base/Preferential-Accretion-of-Protostellar-Binaries/index.html}} together with the tabulated data. Specifically, if two sink particles are available for accretion from a cell the total specific orbital energy is calculated
\begin{equation} \label{eq: total internal energy}
    \varepsilon = \frac{1}{2} \mathbf{v}^2 - {\frac{\textrm{G} (m_{\rm p} + m_{\rm s})}{|\mathbf{r}_{\rm ps}|}}\,,
\end{equation}
where $\mathbf{v}$ is the relative velocity of the sink particles, $m_{\rm p},m_{\rm s}$ are the primary and secondary sink masses, and $\mathbf{r}_{\rm ps}$ is their relative distance. If the system is bound ($\varepsilon < 0$) the eccentricity can then be found
\begin{equation}\label{eq: eccentricity}
    e = \sqrt{1 + \frac{2\ \varepsilon\ h^2}{G (m_1 + m_2)}}\,,
\end{equation}
where $h=\mathbf{r}_{\rm ps} \times \mathbf{v}$ is the specific relative angular momentum. The maximum value in the table lookup for eccentricity is 0.8, and therefore we cap the eccentricity to this, even for cases where the two stars are unbound. Using a table lookup, the fraction of gas accreted to each sink particle can then be determined from $\lambda$.
If two stars have a separation less than the gravitational smoothing length the orbital velocities are smaller, and the derived eccentricity is likely incorrect. This will impact the table lookup and therefore how the total accretion is distributed on to the two sink particles. How large a fraction of the sink particles that are affected by this depends on the size of the gravitational smoothing length. In this paper we have used one third a cell size.

\begin{figure}[t!]
    \centering
    \includegraphics[width=\columnwidth]{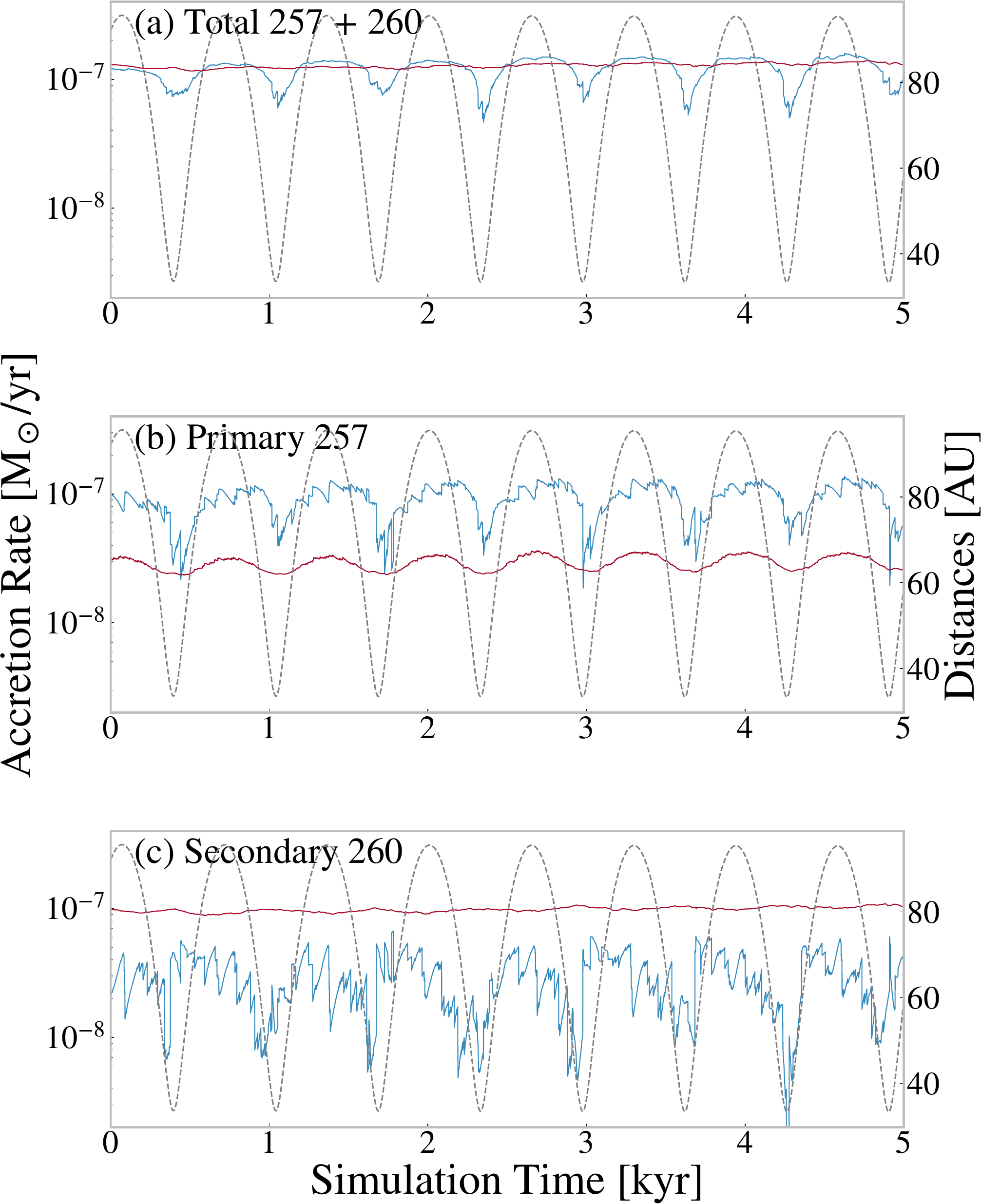}
    \caption{The total and individual accretion rates for the pair 257, 260 at levelmax = 14. In blue is shown the original algorithm, while in red is the evolution with the new algorithm. The sink pair exhibit order of magnitude changes in the individual accretion rates over an orbit with the original algorithm. The distance between the sinks in AU is shown as the dashed line and on the right y-axis. The pair has an average eccentricity of 0.5 and a mass ratio of 0.9.}
    \label{fig: before-after accretion}
\end{figure}

\section{Modelling star formation}
To explore the impact of preferential accretion we use a model of a molecular cloud that was extensively explored and validated by \citet{Haugbolle2018}, and used in previous studies addressing multiplicity statistics \citep{KuruwitaHaugbolle2023}, stellar evolution \citep{Jensen2018} and accretion \citep{Pelkonen+2021}. 
Follow-up simulation with identical setup, but higher resolution also served as the basis for studies addressing the question of water deuteration in stellar clusters \citep{Jensen+2021}, or the role of post-collapse (late) infall in the star formation process \citep{Kuffmeier+2023,Kuffmeier+2024,Pelkonen+2025}.
The physical domain of the molecular cloud is a $L_{\rm box} = 4$ pc periodic box with a total mass of $M_{\rm box} = 3000\ $M$_{\odot}$.
Further, we assume an isothermal sound speed of $c_{\rm s} = 0.18 \ $km s$^{-1}$, which corresponds to $T \approx 10\ $K and a mean molecular weight of $\mu = 2.37$, which is a good approximation for a cold molecular cloud. The mean number density is $n_\textrm{mean} = 795 $cm$^{-3}$. The virial parameter is chosen such that the molecular cloud there is a balance between kinetic energy and gravitational energy, $\alpha_{\rm vir} = 0.83$. The mean magnetic field strength is set to $\rm 7.2 =  \mu $G.
Finally, the dynamical and free-fall times based on these parameters are $t_{\rm dyn} = 1.08 \ $ Myr and $t_{\rm ff} = 1.18 \ $Myr. 
Compared to observations, the adopted conditions are similar to those in the star-forming region of Perseus \citep[][]{Arce+2010,KuruwitaHaugbolle2023}. 

The adaptive mesh is refined according to density and happens when $\rho > \rho_{\rm max}(\ell)$, $\rho_{\rm max}(\ell)$ is the maximum density for a given level $\ell$ and it is given by Eq.~\ref{eq: rho max},
\begin{equation}\label{eq: rho max}
    \rho_{\rm max} = \rho_{\rm levelmin} \left(\frac{\Delta x(\ell)}{\Delta x_{\rm min}} \right)^{-2},
\end{equation}
where $\rho_{\rm levelmin}$ is the density of the minimum level, $\Delta x$ is the cell size, and $\Delta x_{\rm min}$ is the cell size of the minimum level. 
Given the isothermal equation of state, this corresponds to keeping the Jeans length resolved by at least 28 cells at all refinement levels except the highest level, to suppress artificial fragmentation \citep{Truelove1997}.

The maximum level of refinement used to obtain the sample of stars is levelmax$=14$, which corresponds to a cell size of $\Delta x = \frac{4\, \mathrm{pc} }{2^{\rm levelmax}} \approx 50$ AU.
Using this dataset as the basis, we carried out higher resolution simulations, similar to zoom-in simulations \citep[for more details see][]{Kuffmeier+2017}, but applied globally in the box, to study the accretion process for 8 selected sink pairs for stretches of 5 kyr at higher resolution (up to levelmax=18, which corresponds to a resolution of $\Delta x \approx 3$ AU). 

\section{Results} \label{results}
As part of validating the preferential accretion method we have performed a self-consistency test, since the preferential accretion method is based on idealised simulations of black holes and theory about active galactic nuclei.
For star formation there are many interactions in a global simulation, and the gas in the larger-scale environment plays an important role, so it is crucial to test this.
Further, we have performed numerical convergence tests with increasingly higher resolution, until the individual binary systems are resolved.
In this way, we can directly compare the accretion rates for binary stars using the standard recipe and the newly developed algorithm of preferential binary accretion. 
From these simulations we have about 40 binary pairs, and from these we select pairs that have a minimum periastron greater than 0.25 cells and a maximum apastron less than 2.3 cells at levelmax = 14.
This is such that the pairs should be mostly or entirely resolved at levelmax = 18 so preferential accretion is inactive at this higher resolution.
Details of the 8 pairs selected can be found in Table \ref{table:binary table} below. 
The key results are described in the following sections. 
For clarity, we only show and discuss the key results of using the new recipe for the binary pair of sink 257 and sink 260 throughout the paper. 
However, the same plots as shown for the reference binary star in figures 2 and 4 can be found for the other seven pairs in the appendix. 
Note that the sinks are ordered by their creation time, e.g.~sink 257 is the 257th sink particle that formed, and sink 260 the 260th respectively. 

\subsection{Eliminating unphysical suppression of accretion}
The main result is that we can eliminate the unphysical suppression of accretion that was present for most binary systems with small separations in previous simulations; in particular at perisastron
The unphysical accretion was a result of Eq.~\ref{eq: f_v}. 
It created spurious fluctuations in the accretion rates for tight binaries, especially for those that were located inside the same cell. Similar effects must be present in other codes that base accretion efficiency on an energy criteria and / or the relative velocity between the gas cells and the sink particles.
The reason is that the older sink accretes most of the available material, while there is no or only little material left for accretion onto the younger sink (which usually corresponds to the lower-mass secondary star).
We show an example of this in Fig.~\ref{fig: before-after accretion}.
The example is based on the binary pair of sinks 257 and 260, which has a mean distance of 1.4 cells for the maximum level of refinement of $\mathrm {levelmax}=14$ 
in the reference run without activation of the new algorithm. 
Accretion rates for the other seven binary pairs studied (see Table \ref{table:binary table}) can be found in the appendix.
Without the new recipe, 
the sink accretion rate of the primary (sink 257) varies by a factor of a few during each of the almost 8 orbits displayed in the 5 kyr time evolution.
The accretion rate of the secondary (sink 260) even varies by more than a factor of 10 during some orbits.  

In contrast, the accretion rates are much smoother and more regular when the new accretion recipe is applied. 
The top panel corresponds to the total accretion rate and it shows how the preferential accretion algorithm keeps the total accretion close to the same as in the reference run, but without the spurious dips in the accretion rate at periastron during the orbits.
The accretion rate of the primary is systematically lower (middle panel), the one of the secondary systematically higher (lower panel) compared to the reference run without the new recipe.
This is a consequence of the underlying idealised scenarios of the binary accretion models that are used to derive the values of $\lambda(q)$. 
In these scenarios, the lower mass object (i.e., the secondary) moves through the denser gas and thereby accretes more material when it orbits around the more massive object (i.e., the primary).
This effect is most pronounced for the binary pairs with low $q$, i.e., sink pairs with high difference in mass between primary and secondary. 
For the cases of binaries with objects of similar mass (sink pairs 75/82, 142/144, 217/219, 253/254) this effect is less pronounced (Fig.~\ref{fig: accretion-A}). 
In fact for the almost equal mass binary (253/254), the overall accretion rates of the primary and the secondary are almost identical to the runs without the recipe, except for the important improvement of quenching the spurious accretion dips at periastron.
 With the algorithm active we observe that the accretion rate only shows very little variability during the time interval of 5 kyr, which matches what we see for accretion rates of single sinks in otherwise similar regions.
 It is worth noting that in this example the sinks are located within 0.7 and 1.9 cells of each other during the time interval of 5 kyr.
 Increasing the resolution by a few levels of refinement eliminates the unphysical suppression of accretion also without activation of our new recipe as shown in the following subsection.

\subsection{Resolution study}
The main purpose of this resolution study is to make sure that the preferential accretion algorithm performs as intended at all resolutions where the binary pairs are not resolved, and to gain a better understanding of the limitations of the algorithm.
First we inspect the accretion rates in Fig.~\ref{fig: 257-260 accretion}, where we find that the accretion rate for higher resolution is higher for primary sinks, but the reverse is true for secondary sinks, where the lowest resolution run has the highest accretion rate.
The higher accretion rate for the primary and lower accretion rates for the secondary in the high resolution runs compared to the low resolution runs where the recipe is active is a consequence of the discrepancy between the accretion dynamics resolved in the zoom-in simulations and the assumption of an isolated circumbinary disk reservoir. 
Fig.~\ref{fig: pairwise preferential accretion with data} shows the significant variations of $\lambda$, especially for values of $q<0.5$. 
On purpose, we show the case of sink 257/260 that has the most significant differences in $\lambda$ for various resolution. 
The example clearly illustrates the caveat of the preferential accretion algorithm.
It is only as good as the dynamics of the specific protostellar pair match the idealised scenario that was used to compute the table values. 
For this work, we are adopting values for $\lambda$ that were computed based on an idealised scenario of binary accretion from a circumbinary disk. 
However, the protostars are embedded in turbulent molecular cloud environments with more complicated overall dynamics (Fig.~\ref{fig:Sigma_map}). 
The resolution study reveals that the primary (secondary) sink accretes more (less) material than in the idealised scenario of accretion from an isolated circumbinary disk,
the primary star accretes systematically less material than for younger binary systems, especially for significant mass differences between the two binary components.

Apart from that, the lower resolution runs have a smoother profile total accretion rate with less dependence of the distance, which matches the expectations of more complexity when the system is more resolved.
However, the accretion rates vary by much less than an order of magnitude, which means we get a good approximation without resolving the system.
There is an interesting observation for the levelmax = 16 run, which is that near periastron it aligns almost entirely with the accretion rates at lower resolutions, but otherwise it aligns with the higher resolutions.
This is because at this resolution the binary is resolved partially, such that for most of the time their accretion radii are not overlapping enough to activate the binary accretion recipe, but near periastron they overlap enough for the binary accretion recipe to dominate the accretion.

\begin{figure}[t!]
    \centering
    \includegraphics[width=0.86\columnwidth]{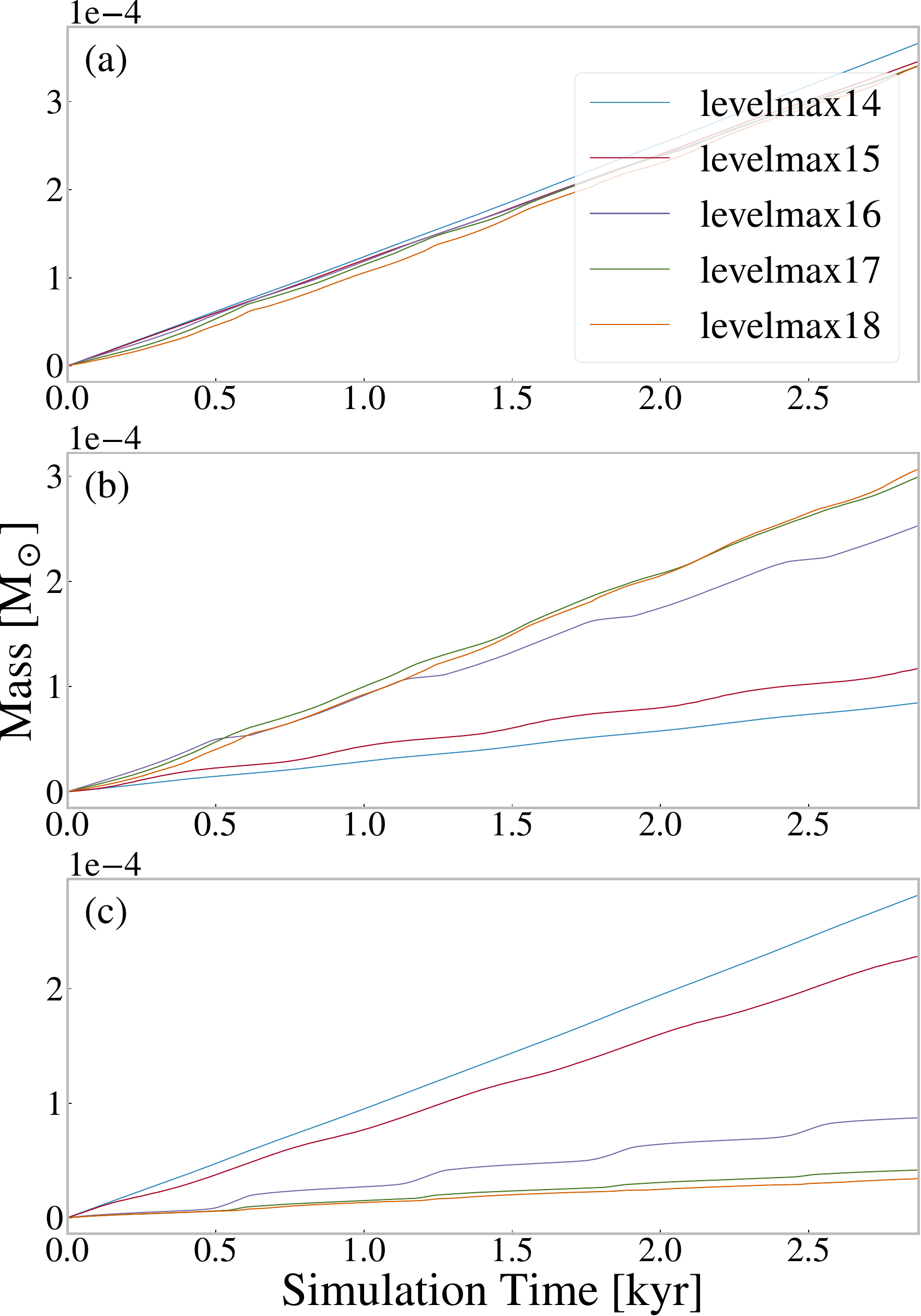}
    \caption{The total (a), primary (b), and secondary (c) integrated masses of the sink pair 257, 260 over different resolutions from levelmax = 14 to 18. The origin is set to where the levelmax = 18 run started.}
    \label{fig: 257-260 integrated mass}
\end{figure}

\begin{figure}[t!]
    \centering
    \includegraphics[width=\columnwidth]{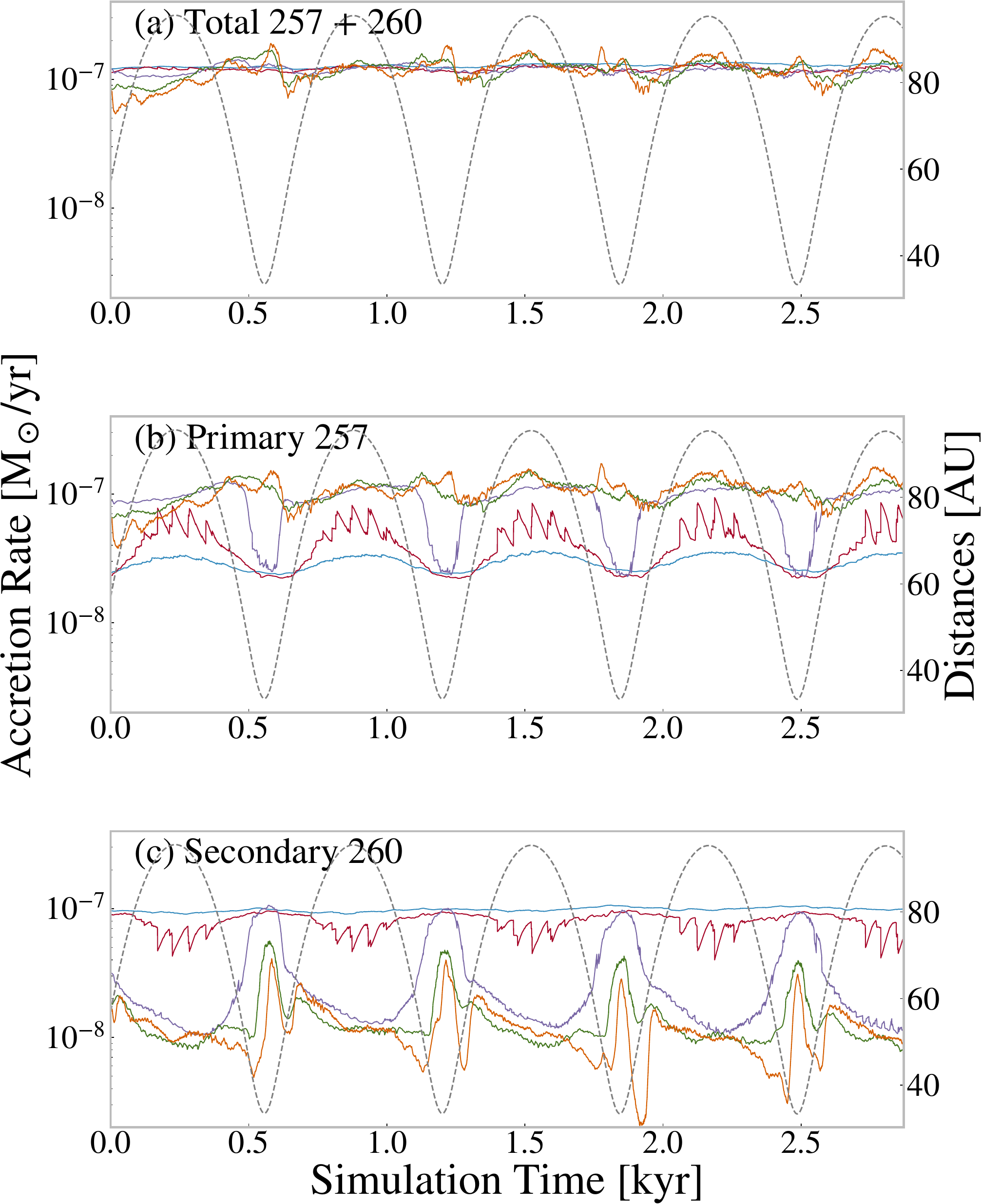}
    \caption{The total (a), primary (b), and secondary (c) accretion rates of pair 257, 260 over different resolutions from levelmax = 14 to 18. The distance between the sinks in AU is shown on the right y-axis. The pair has an average eccentricity of 0.5 and a mass ratio of 0.9.}
    \label{fig: 257-260 accretion}
\end{figure}

Looking at the integrated mass of this system of the same time-range in Fig.~\ref{fig: 257-260 integrated mass}, we see a similar behaviour for the primary and secondary sinks.
For the total integrated mass we clearly see that the accretion rate is in fact similar for this system regardless whether preferential accretion is active or the system is gradually being resolved.
This is less clear for several of the other systems, where there is some variation. This is likely due to the complexity and turbulence of the surroundings of the binary systems, and this aspect is explored further in the following section. 

\subsection{Idealised algorithm in a complex and turbulent environment}
In Fig.~\ref{fig: pairwise preferential accretion with data} the average $\rm \lambda$ is shown to compare the simulation results to the theoretical plot in Fig.~\ref{fig: preferential accretion}. There is a big spread and only some points agree with the theoretical model. This is an important limitation; the algorithm is not perfect and in real and complex situations there are other aspects affecting the accretion.
Independent of the resolution, Fig.~\ref{fig:Sigma_map} shows the complexity of the turbulent environment. In the complementary figure \ref{fig: pairwise preferential accretion with data} the average $\rm \lambda$ without the theoretical curve is shown to visualise each pair and the run it is from. Some pairs have a big spread in $\rm \lambda$ like 257, 260 that we compared above, but others like 75, 82 and 292, 290 have a much smaller spread.
This illustrates that all the selected pairs have different characteristics within the constraints set, and this also means that we cannot expect exact similarity.

\section{Discussion}
\label{discussion}
We have successfully managed to eliminate the unphysical suppression of accretion of unresolved close binaries, which was the main problem with the previous accretion approaches.
This is seen across all the figures in the resolution study above, in particular in Fig.~\ref{fig: before-after accretion}, where we observe that the maximum difference is 2 times the minimum value, and for the most part the variability is less than about 0.4, which is a big improvement compared to not having a binary sub-sink accretion algorithm.
We find that the variability is larger for levelmax = 17 and 18, which is due to a more complicated environment around the binary system as shown in Fig.~\ref{fig:Sigma_map}, where e.g.~the individual discs and outflows begin to be marginally resolved.

\begin{figure}
    \centering
    \includegraphics[width=\columnwidth]{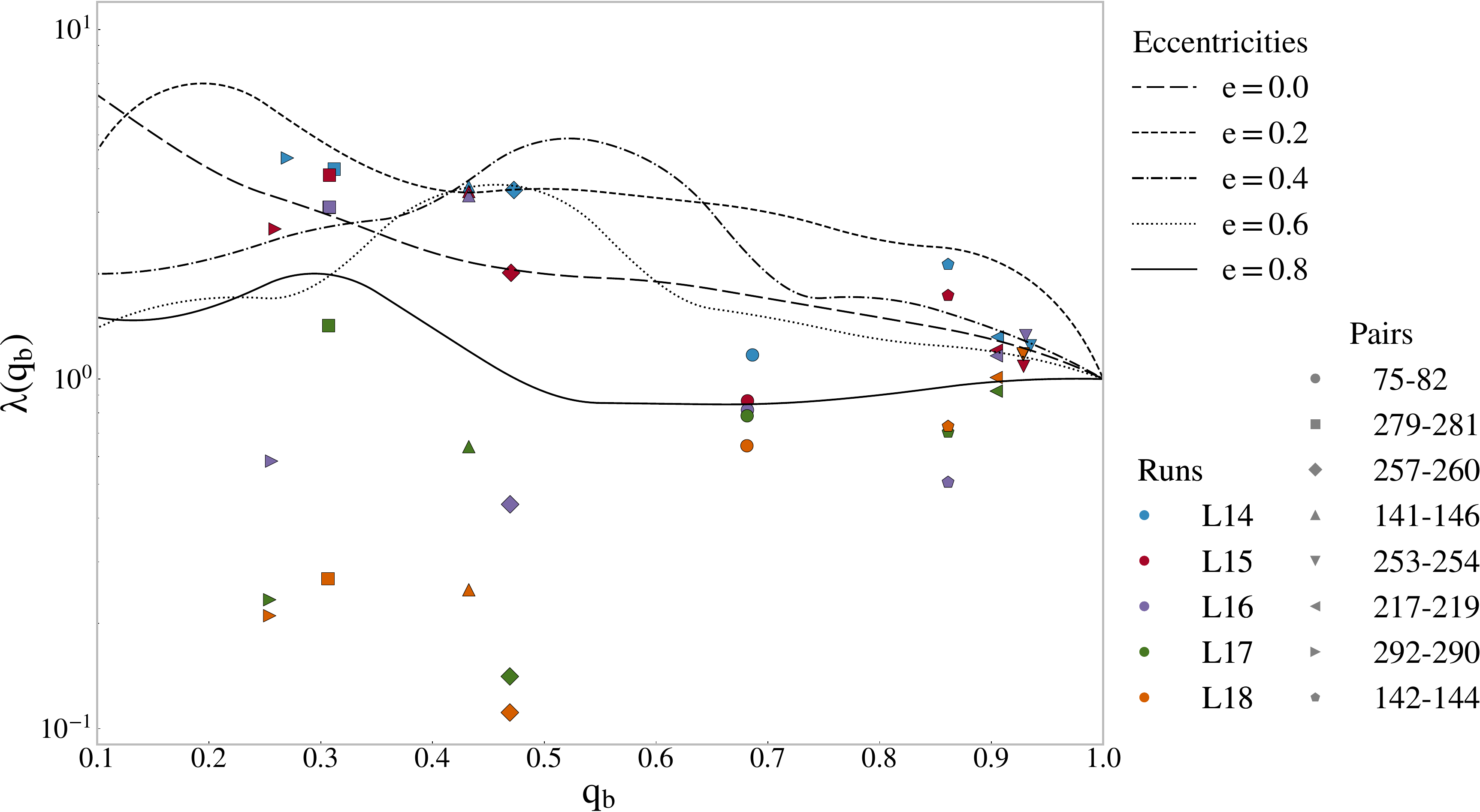}
    \caption{The average $\lambda$ of each pair in each run against the mass ratio. The colour symbolises the resolution of the run it comes from and the symbol which pair it is. Lines are from Fig.~\ref{fig: preferential accretion}.}
    \label{fig: pairwise preferential accretion with data}
\end{figure}

The systematic offset in the accretion variability seen for the low-resolution runs without activation of the new sink algorithm has a large impact. 
Most simulation codes use a similar method of accretion without adopting a recipe for preferential accretion and thereby experience unphysical suppression of accretion for tight binaries.
We show that this artificial quenching of accretion can be eliminated with our new approach using a simple implementation of an idealised framework for accretion onto binary black holes and active galactic nuclei.
It means that for future simulations it is possible to carry out global runs, while still obtaining reliable accretion rates for binary systems at lower resolution than without using the new algorithm.
This is helpful for investigating binaries in large-scale simulations because we can carry out larger and/or longer simulations at lower costs, which allows us to get a better understanding of the accretion process in entire star-forming regions.
We find more even accretion rate curves for lower resolution simulations, while the average accretion rates are very close to each other.
The lower resolution simulations show less variability and therefore more flat accretion profiles.
This is likely due to the fact that when increasing the resolution, the surroundings become more resolved and we also start to resolve more complex structures such as discs and outflows, some of which can be seen in Fig.~\ref{fig:Sigma_map} as well as in the high resolution column density maps of the other binary pairs shown in Fig.~\ref{fig:Sigma_map_075} to Fig.~\ref{fig:Sigma_map_290} in the appendix.
These can significantly affect the accretion process \citep[see review by][and references therein]{Tsukamoto2023} including the launching of outflows \citep{Podio2021}.

In a more detailed account, the mass fractions accreted by the sink particles, seems to be well converged for mass ratios close to 1. This is particularly visible in Fig.~\ref{fig: pairwise preferential accretion with data}, where for the two highest mass ratio pairs we see good agreement in $\rm \lambda$ across resolutions.
This also matches the expectations that for a mass ratio approaching 1, $\rm \lambda$ approaches 1.
We also find that there is a large spread in $\rm \lambda$ for lower mass ratios. Specifically we see that the primary star accretes significantly more than the secondary in the realm where the idealised model predicts the secondary star to accrete significantly more.
However, these cases happen at higher resolution when the binary accretion recipe is not active or only partially active.
The key takeaway from this result is that there is little understanding of binary accretion for low mass ratios $q$, and more work is required to understand this regime for proto-binaries in a gas-rich environment.

In the current paper, we are limited to 8 binary systems and cannot cover the full parameter space to better characterise the spread of $\rm \lambda$ against mass ratio and eccentricity, but limit ourselves to a first presentation of the new preferential binary accretion recipe. The pairs are specifically picked for being different to each other, such that we can study the implications over the widest range of systems possible in our simulation data.
For future research,
it is worth investigating and better understanding the impact of eccentricity on the accretion process of binary systems \citep[such as studied by][]{Oneill2024} in the context of star formation in turbulent star-forming regions. 

\section{Conclusion}
In this paper we addressed accretion in unresolved binary systems. Existing accretion recipes introduces high time variability and have assumptions that cannot capture the physical flow adequately, such as preferential accretion to the secondary star in the system.
To overcome spurious accretion in binary systems, we introduce a new recipe for preferential binary accretion to handle the accretion in binary systems, which is based on theoretical models from active galactic nuclei and idealised simulations of binary black holes. By implementing this in {\sc ramses} we attempt to eliminate this unphysical accretion and obtain a better approximation of accretion for unresolved binary systems.  

The new accretion recipe successfully eliminates the unphysical dips in accretion rates at periastron that occur in the version without using the recipe.
This is a significant improvement over current models and can easily be extended to other codes than {\sc ramses}, which will allow us and others to obtain a better description of binary systems accretion in global simulations.
In addition, we track the fraction of mass available for accretion for the binary components. This is particularly useful for close binaries, often in the same cell at lower resolutions, and it allows us to give a good approximation to the accretion for mass ratios close to 1. The combination of these enables the option to run global simulations at a lower resolution than otherwise required for binaries to provide usable data. 

Although the algorithm proves to be successful, it has limitations.
It is not yet clear how eccentricity and mass ratio affect the accretion for binary systems at lower mass ratios (of order 0.1), however, we have presented the first results of what they look like with the preferential accretion recipe.
Our study shows that substantial differences can occur when adopting the accretion parameters obtained from models of binary accretion from isolated circumbinary disks for sink pairs that are in reality embedded and effected by the turbulent molecular cloud environment. 
In particular, a resolution study shows that the accretion rate of the secondary tends to be overestimated using the table values from isolated accretion models in comparison to models where the embedded binary system is resolved.
We suggest that obtaining accretion histories for more embedded binary sinks in molecular clouds could provide a more refined table of the adopted accretion parameters and thereby optimize the use of the recipe.
More work is needed to determine whether how the recipe can be improved in this regime.
Despite the limitations, this paper presents a promising step forward in modelling close binary systems, in particular in large scale simulations and for better constraints on the accretion histories of embedded binaries in the future. By offering a more accurate and computationally efficient method, we contribute to the broader effort to understand and model star-forming regions and protostellar binaries.

\subsection{Data availability}
A Fortran implementation of the binary accretion module and figures from this paper, and additional column density images at different resolutions, that illustrate the physical and numerical processes are available on the University of Copenhagen Electronic Research Data Archive (ERDA) through the link\footnote{\url{https://erda.dk/vgrid/Preferential-Accretion-of-Protostellar-Binaries}}. The bulk data generated and analysed in this paper are not publicly available due to the large data volume and the ongoing research of other aspects of the datasets, but they are available on reasonable request.

\begin{acknowledgements}
The Tycho HPC facility at the University of Copenhagen, supported by research grants from the Carlsberg, Novo, and Villum foundations, was used for carrying out the simulations, the analysis, and the long-term storage of the results.
MK acknowledges funding from a Carlsberg Reintegration Fellowship (CF22-1014). TH acknowledges funding from the Independent Research Fund Denmark through grant No. DFF 4283-00305B.
\end{acknowledgements}

\begin{figure*}[htb!]
    \centering
    \includegraphics[width=0.82\textwidth,clip,trim=0 0 0 0]{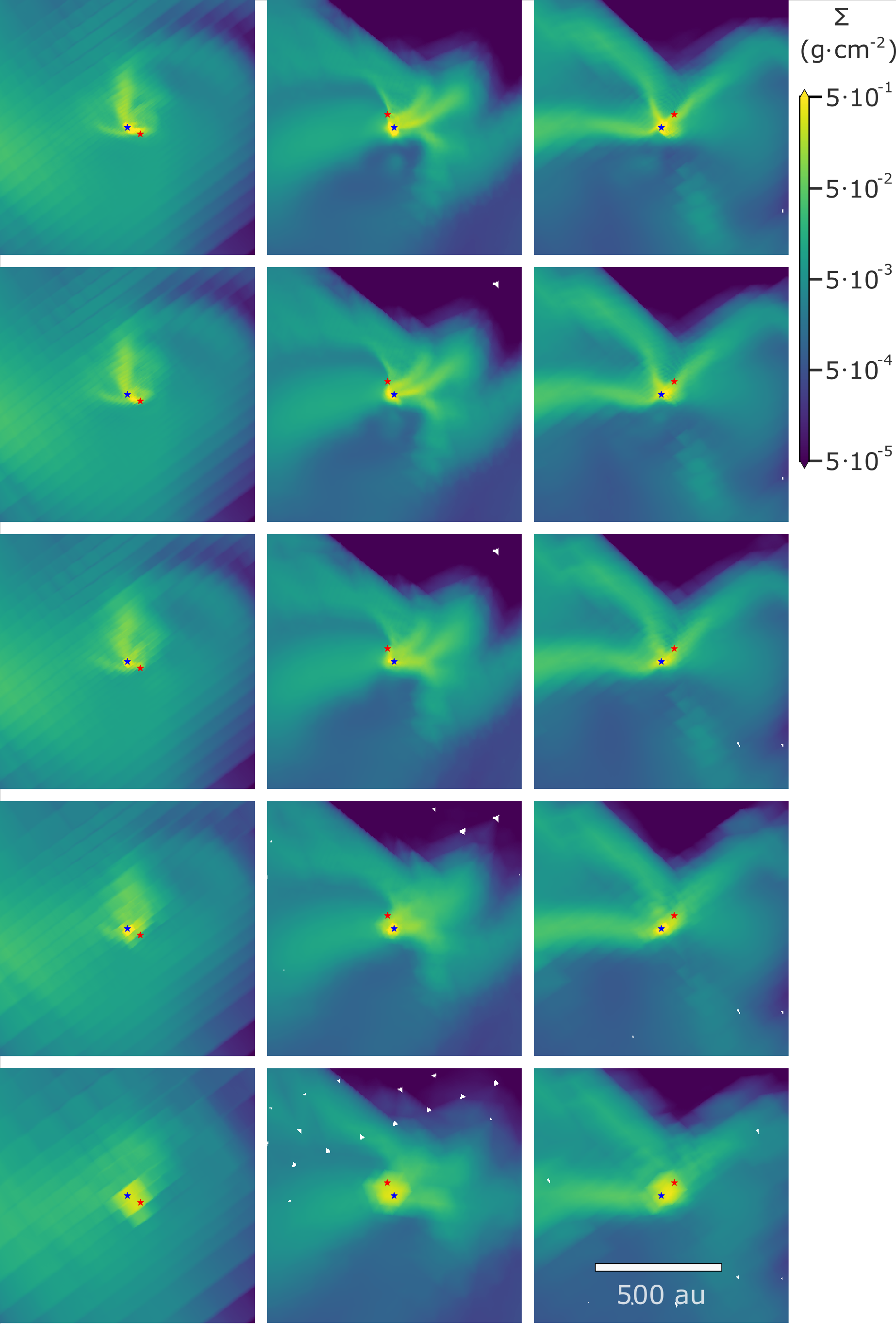}
    \caption{Column density plot on the 3 axes of the primary sink particle in a 1000 AU by 1000 AU field of view, where the primary sink is the blue star and the secondary is the red star. The resolution goes from levelmax = 14 at the bottom panel to levelmax = 18 at the top in steps of 1. It shows the complex surroundings of pair 257, 260 and how with increasing resolution the gas flow to the individual sinks becomes mostly resolved.}
    \label{fig:Sigma_map}
\end{figure*}

\bibliography{references}
\bibliographystyle{aa}

\begin{appendix}
\section{Evolution of binary pairs}
\begin{figure*}[htb!]
    \centering     \includegraphics[width=0.82\textwidth,clip,trim=0 0 0 0]{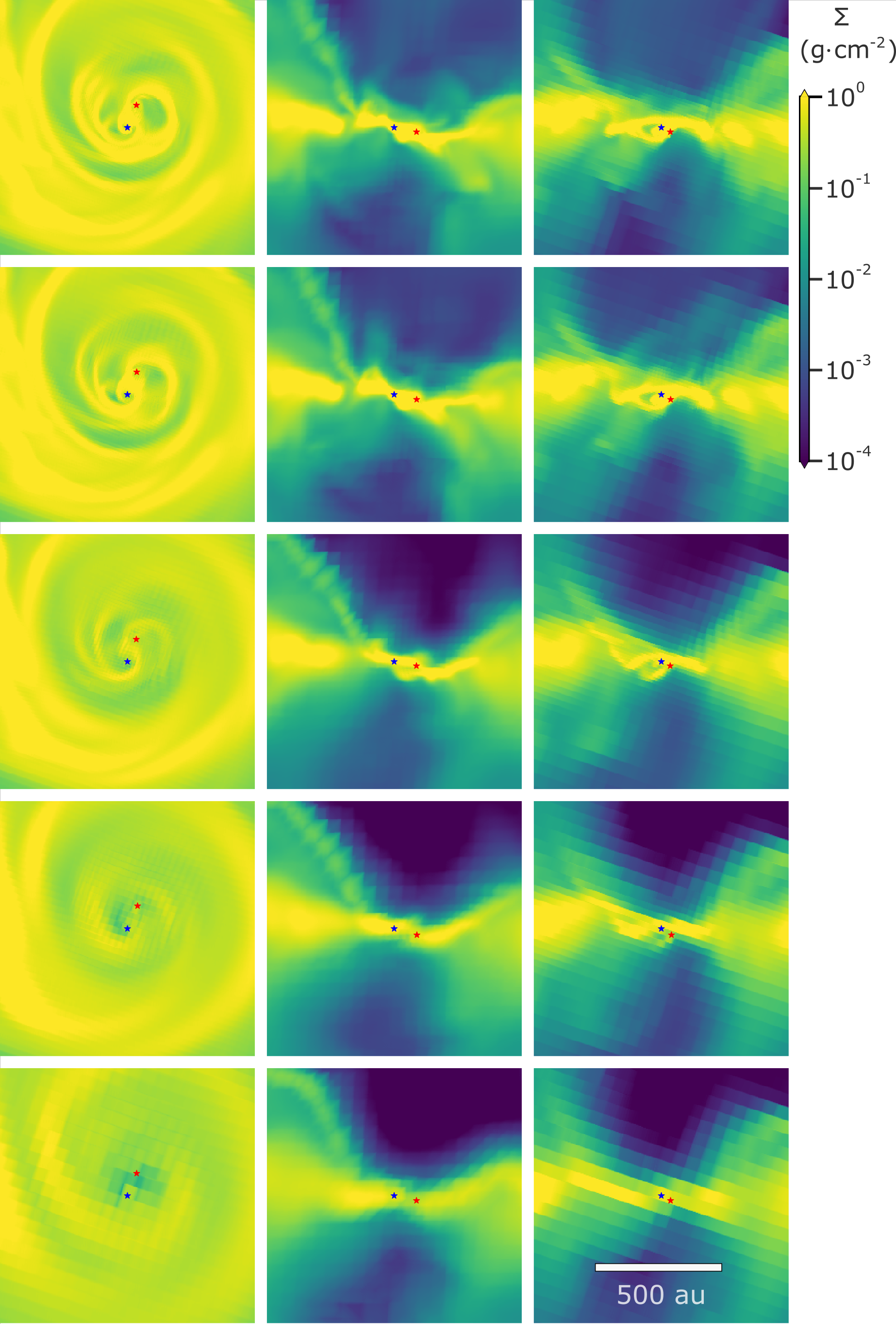}
    \caption{Same as Fig.~\ref{fig:Sigma_map}, but for pair 75, 82. }
    \label{fig:Sigma_map_075}
\end{figure*}

\begin{figure*}[htb!]
    \centering     \includegraphics[width=0.82\textwidth,clip,trim=0 0 0 0]{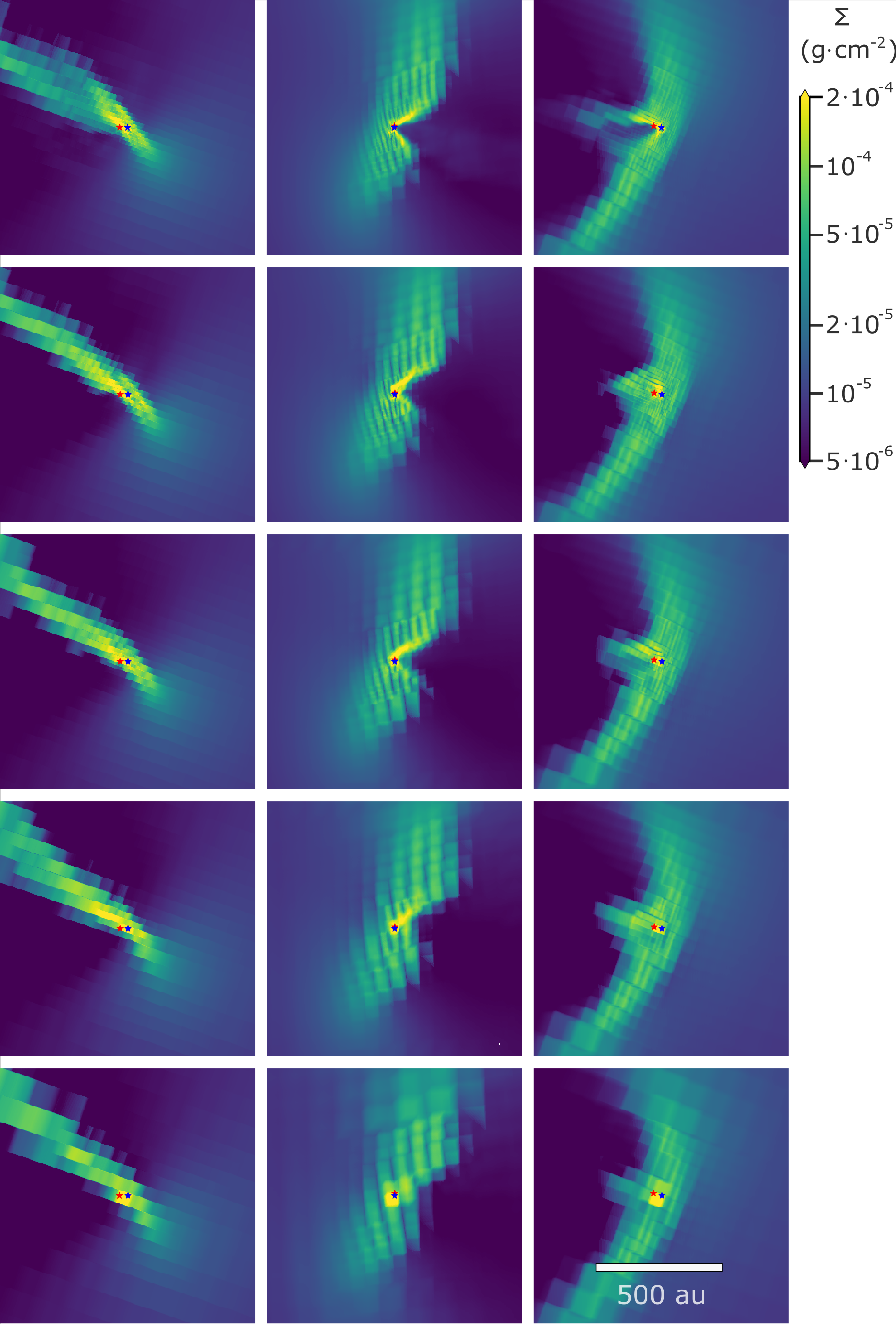}
    \caption{Same as Fig.~\ref{fig:Sigma_map}, but for pair 141, 146.}
    \label{fig:Sigma_map_141}

\end{figure*}

\begin{figure*}[htb!]
    \centering     \includegraphics[width=0.82\textwidth,clip,trim=0 0 0 0]{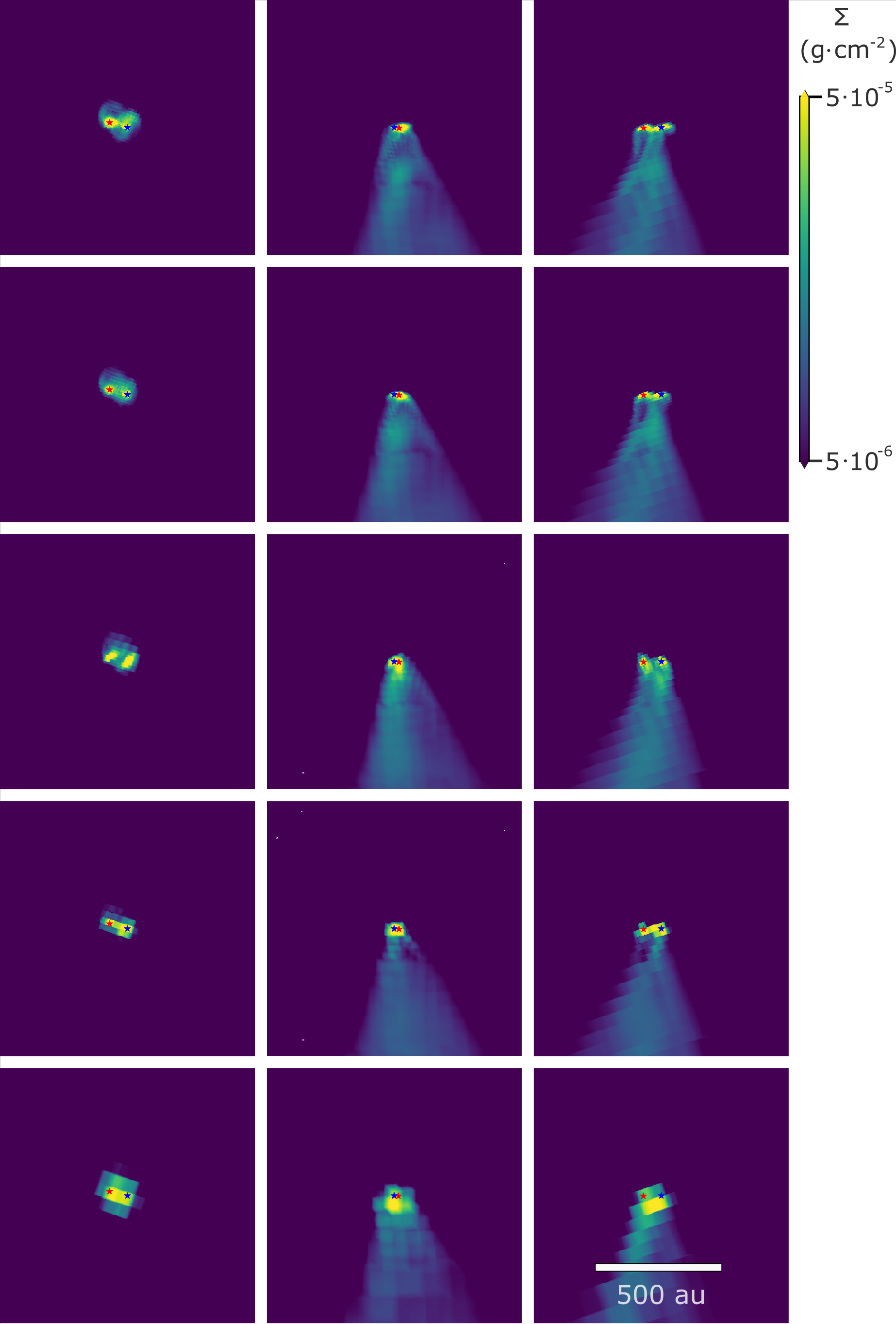}
    \caption{Same as Fig.~\ref{fig:Sigma_map}, but for pair 142, 144.}
    \label{fig:Sigma_map_142}

\end{figure*}

\begin{figure*}[htb!]
    \centering     \includegraphics[width=0.82\textwidth,clip,trim=0 0 0 0]{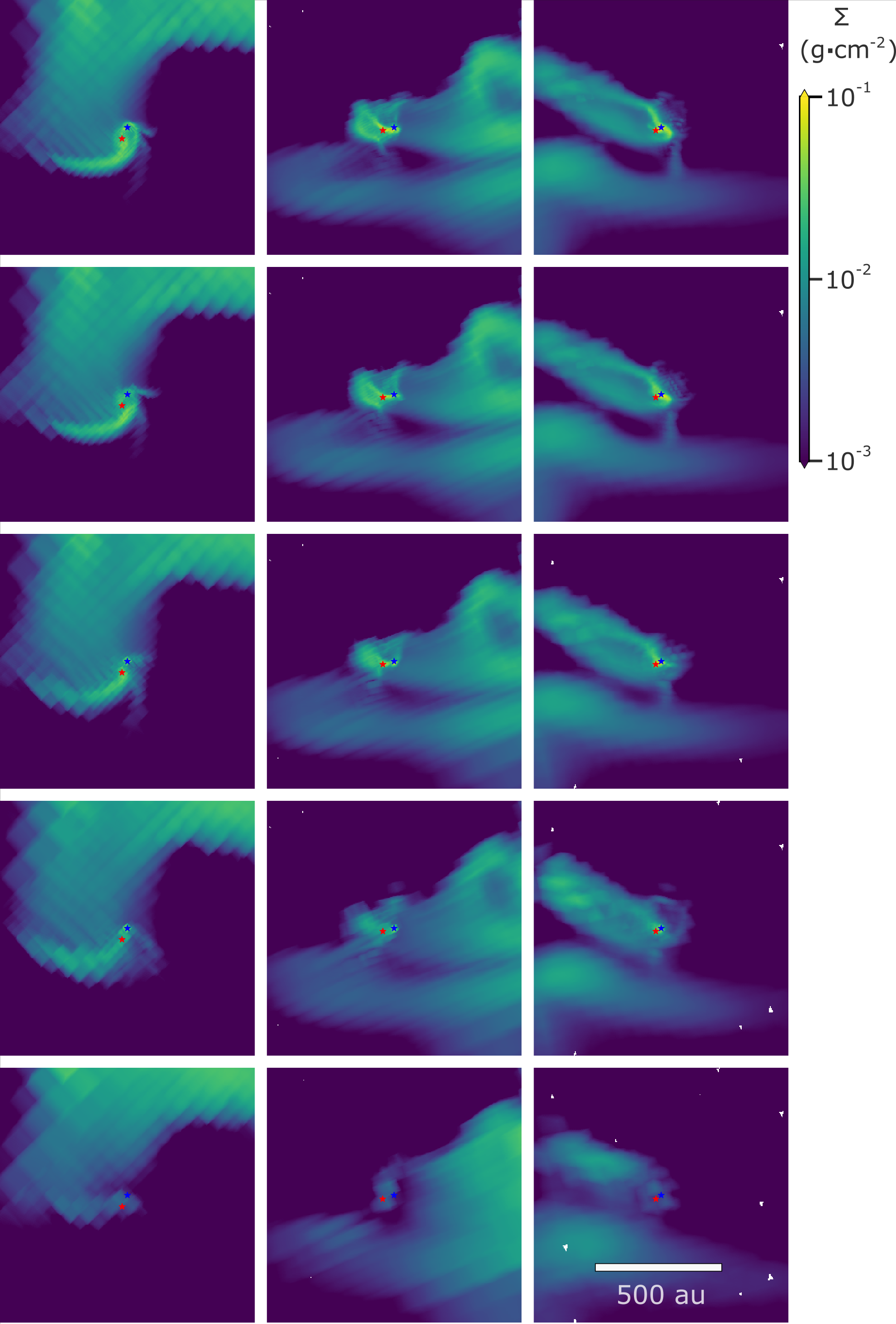}
    \caption{Same as Fig.~\ref{fig:Sigma_map}, but for pair 217, 219.}
    \label{fig:Sigma_map_217}

\end{figure*}

\begin{figure*}[htb!]
    \centering     \includegraphics[width=0.82\textwidth,clip,trim=0 0 0 0]{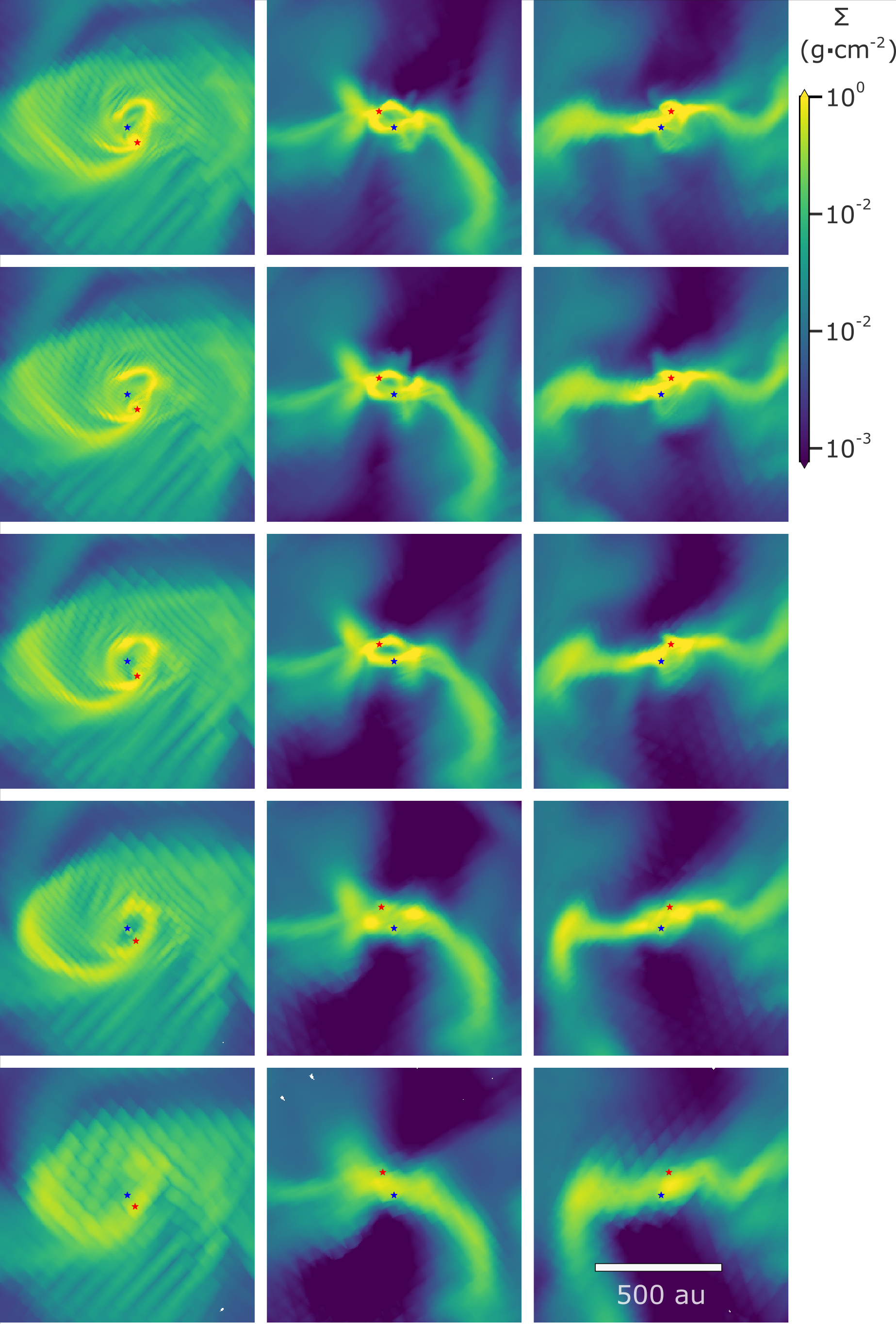}
    \caption{Same as Fig.~\ref{fig:Sigma_map}, but for pair 253, 254.}
    \label{fig:Sigma_map_253}

\end{figure*}

\begin{figure*}[htb!]
    \centering     \includegraphics[width=0.82\textwidth,clip,trim=0 0 0 0]{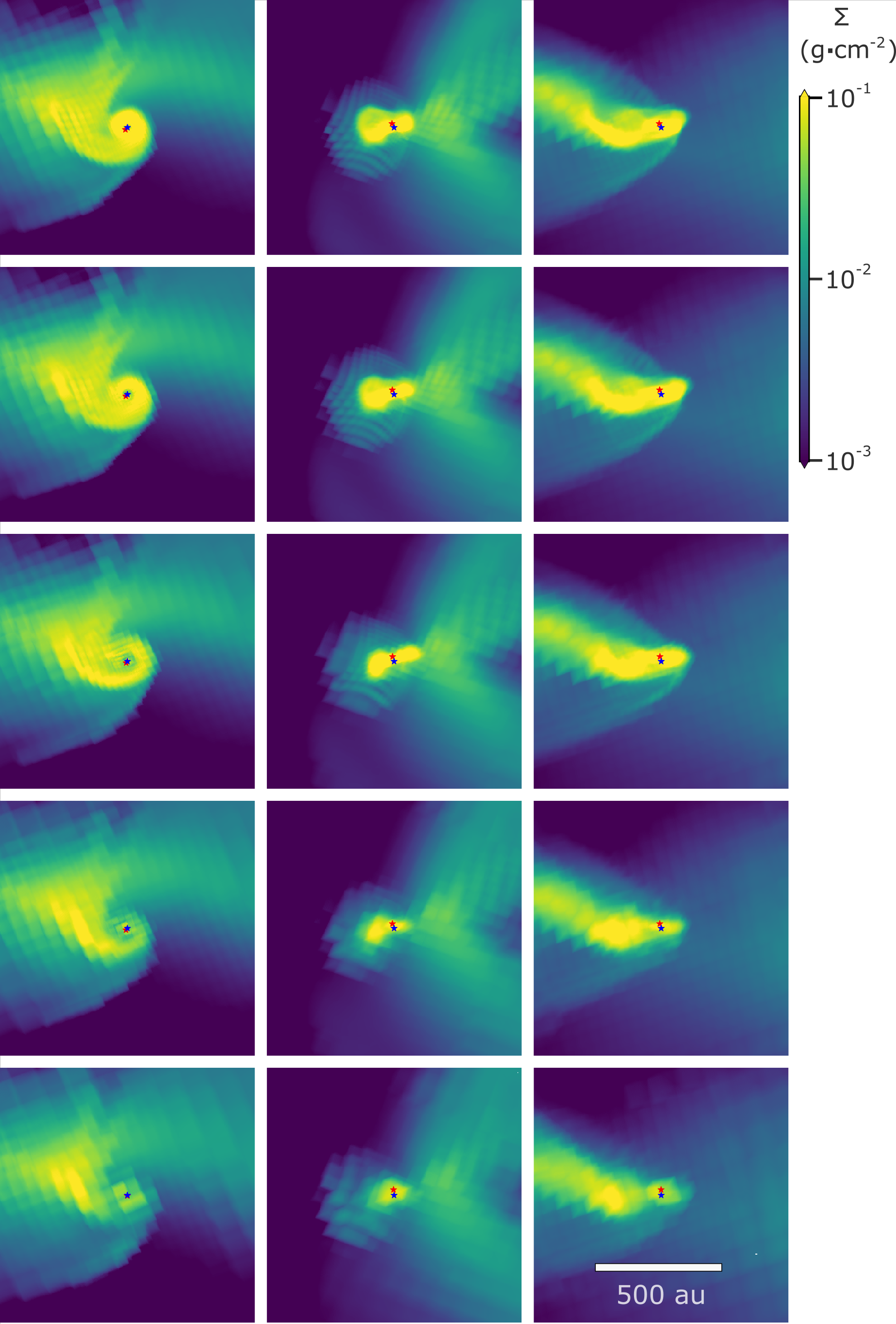}
    \caption{Same as Fig.~\ref{fig:Sigma_map}, but for pair 279, 219.}
    \label{fig:Sigma_map_279}

\end{figure*}

\begin{figure*}[htb!]
    \centering     \includegraphics[width=0.82\textwidth,clip,trim=0 0 0 0]{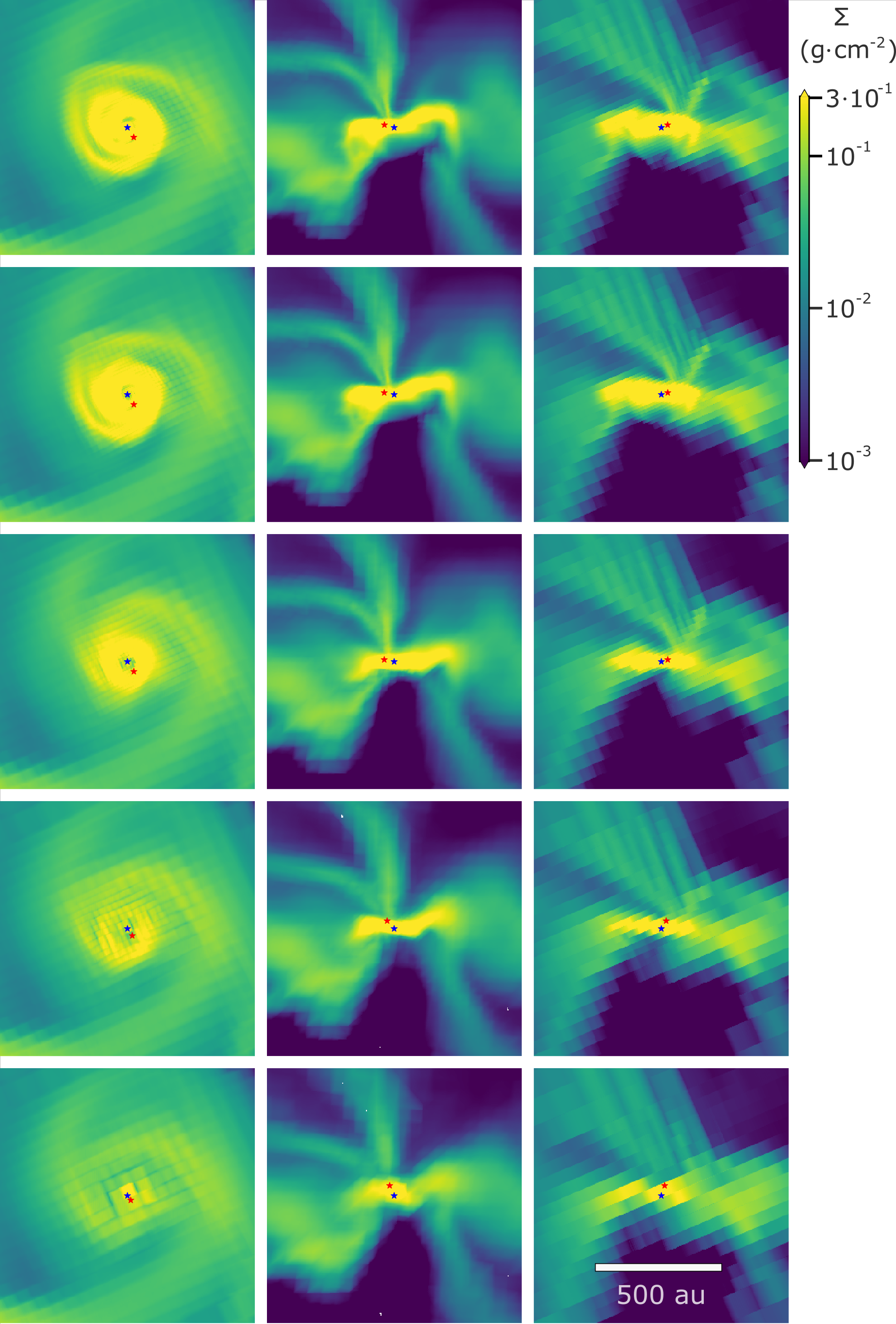}
    \caption{Same as Fig.~\ref{fig:Sigma_map}, but for pair 290, 292.}
    \label{fig:Sigma_map_290}

\end{figure*}

To demonstrate the different accretion profiles of the binary pairs, we show the accretion rates for the other six pairs without and with the implementation of our new algorithm for the different resolutions in Figs.~\ref{fig: before-after accretion-A} and \ref{fig: accretion-A}. The figures are complementary to the ones shown for the reference sink pair 257/260. 
Similarly, we illustrate the various protostellar environments in the column density maps of the binary systems in Fig.~\ref{fig:Sigma_map_075} to Fig.~\ref{fig:Sigma_map_290}, complementary to the column density map shown for the reference sink pair 257/260 in Fig.~\ref{fig:Sigma_map}.

\vspace*{-0.18\textheight}
\begin{minipage}{\textwidth}
\begin{center}
    \hspace{0.635\columnwidth}
    \includegraphics[width=0.3\columnwidth]{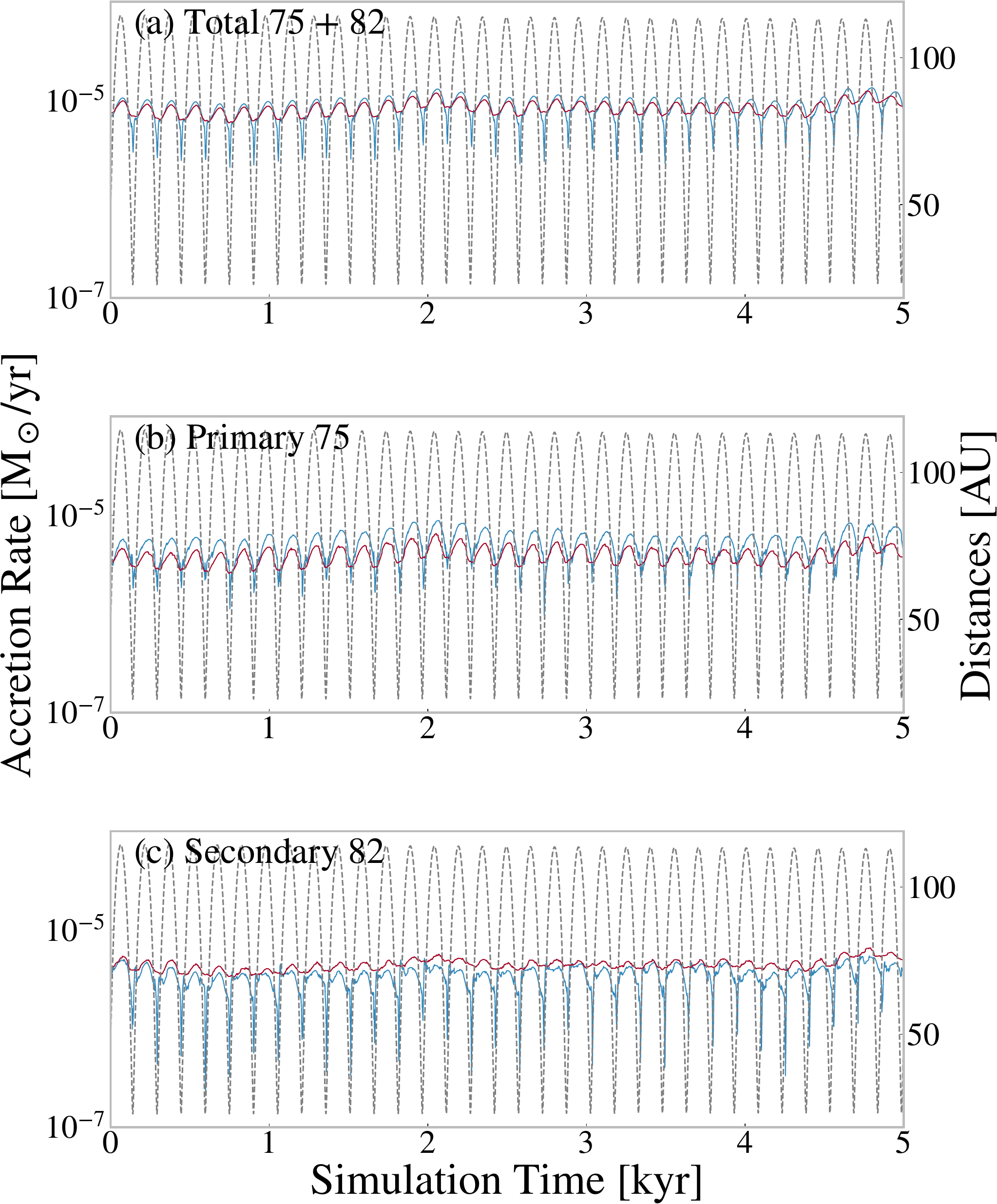}
    \vspace*{0.02\textheight}

    \includegraphics[width=0.3\columnwidth]{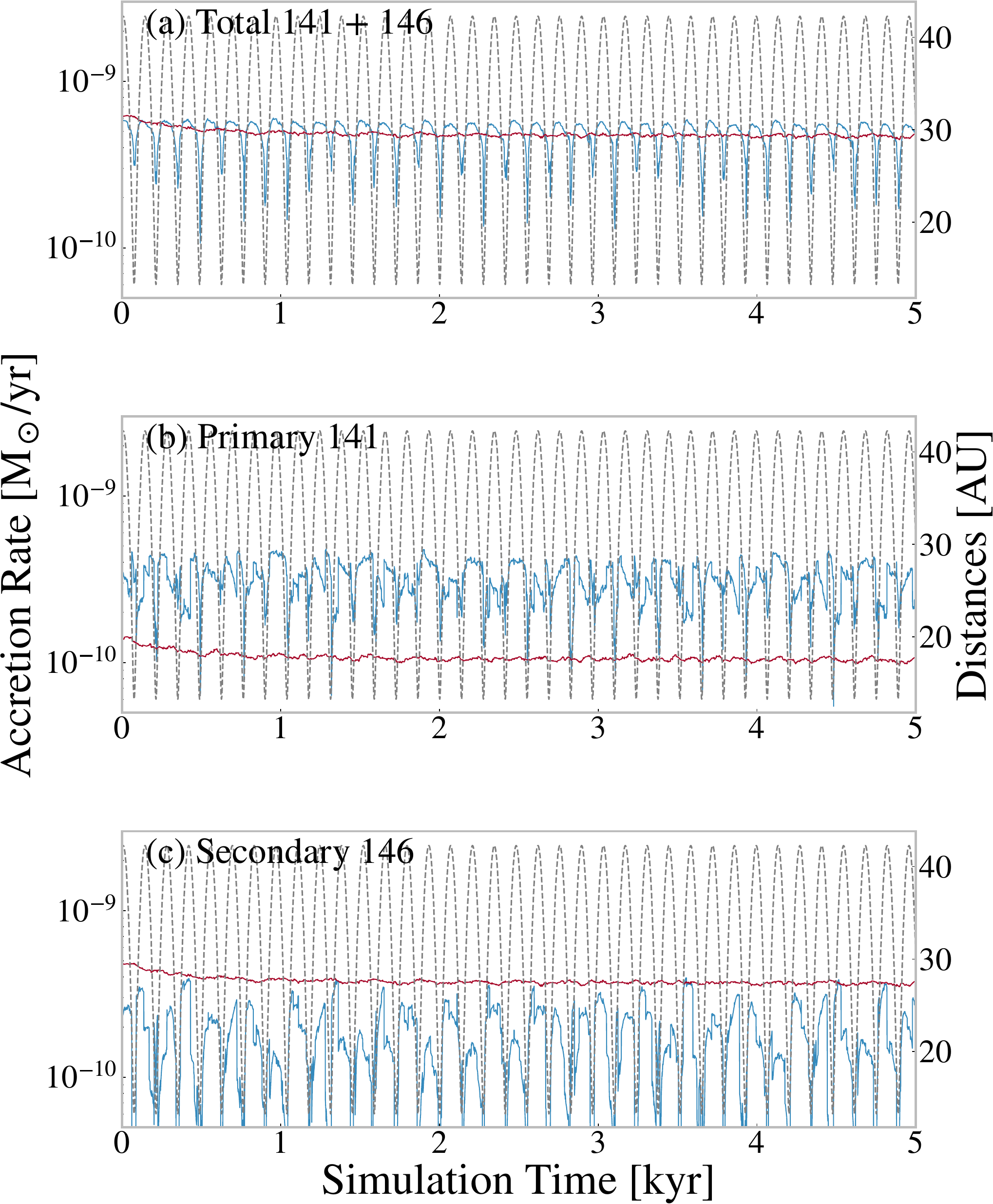}
    \hspace{0.01\columnwidth}
    \includegraphics[width=0.3\columnwidth]{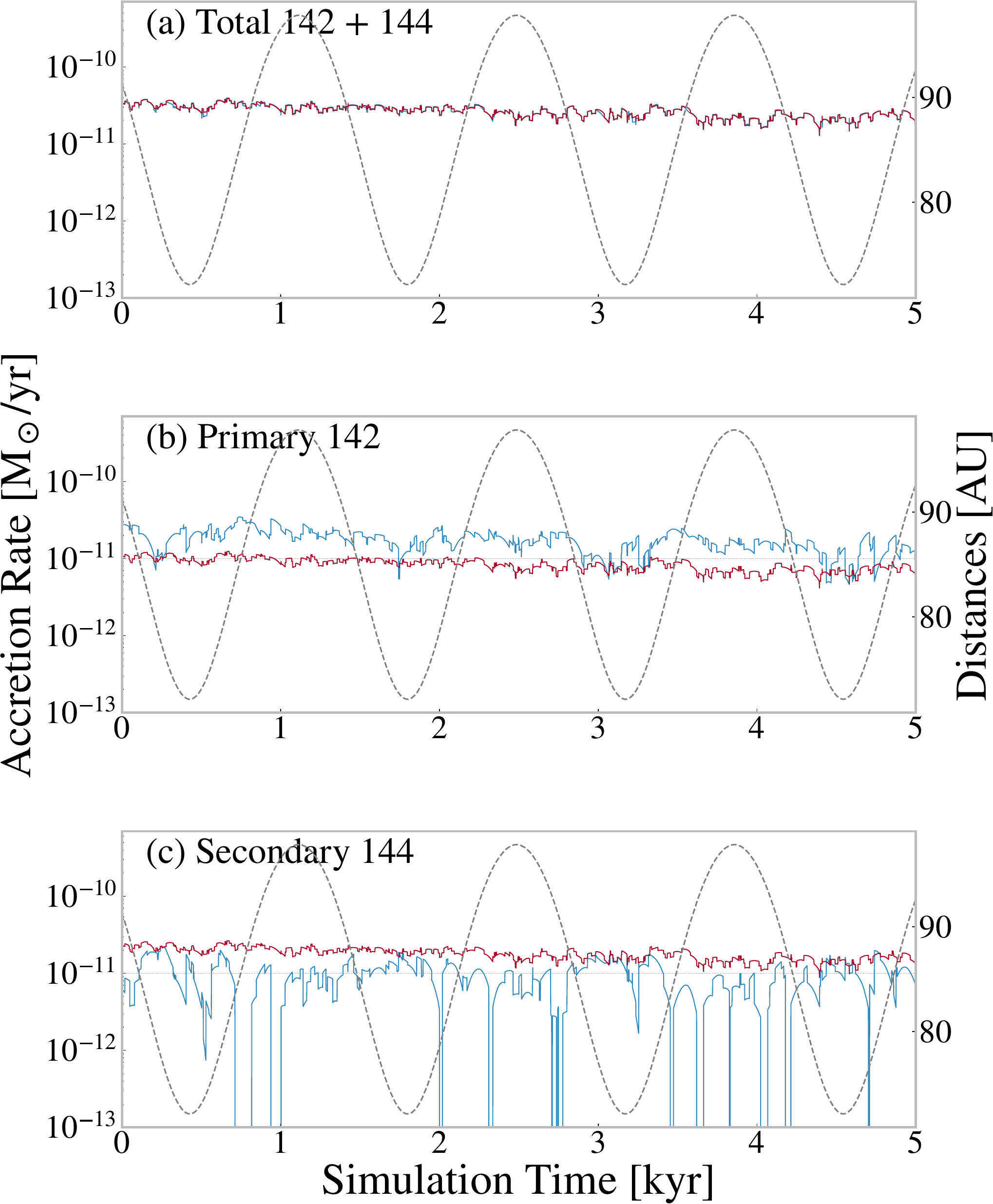}
    \hspace{0.01\columnwidth}
    \includegraphics[width=0.3\columnwidth]{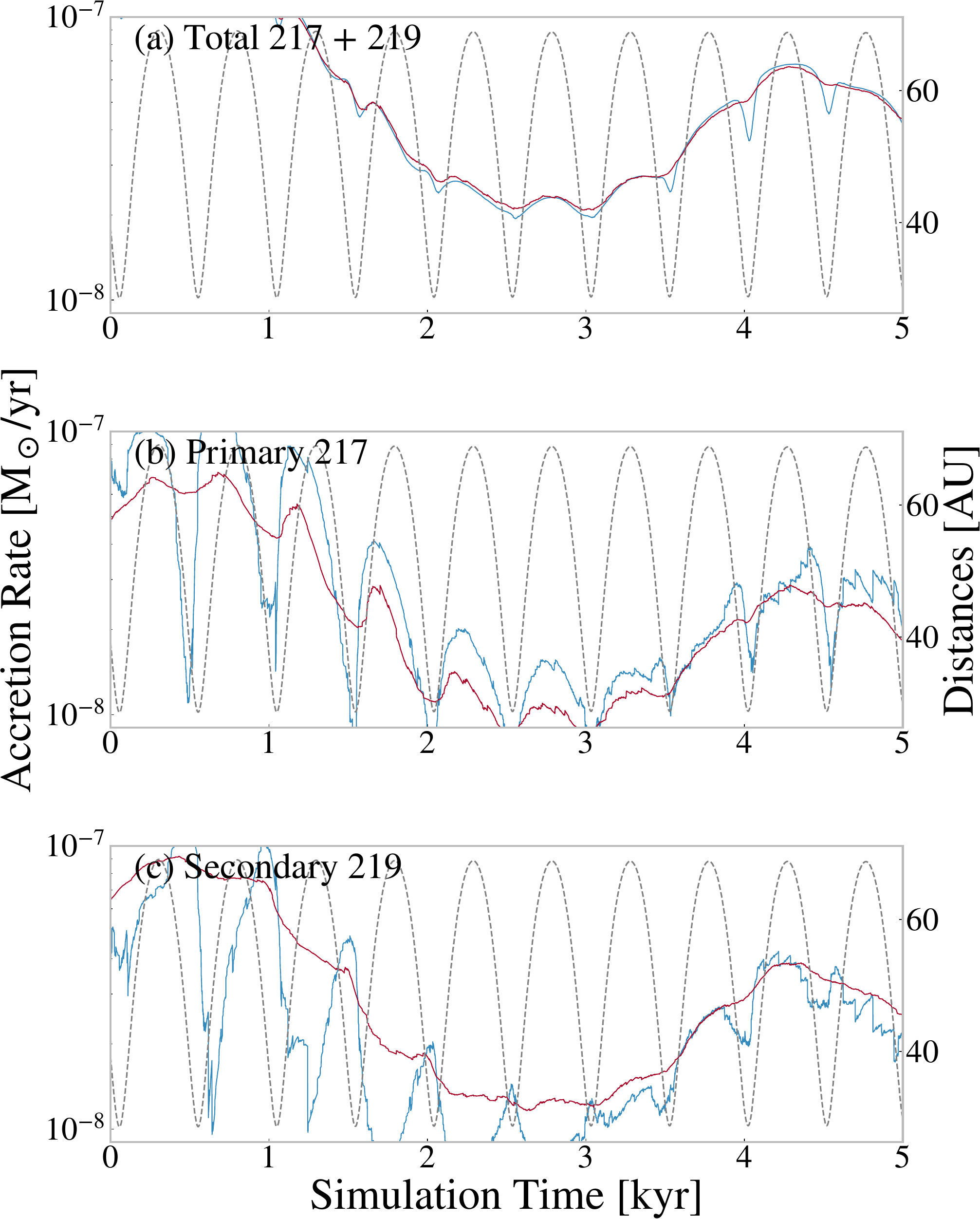}
    \vspace*{0.02\textheight}

    \includegraphics[width=0.3\columnwidth]{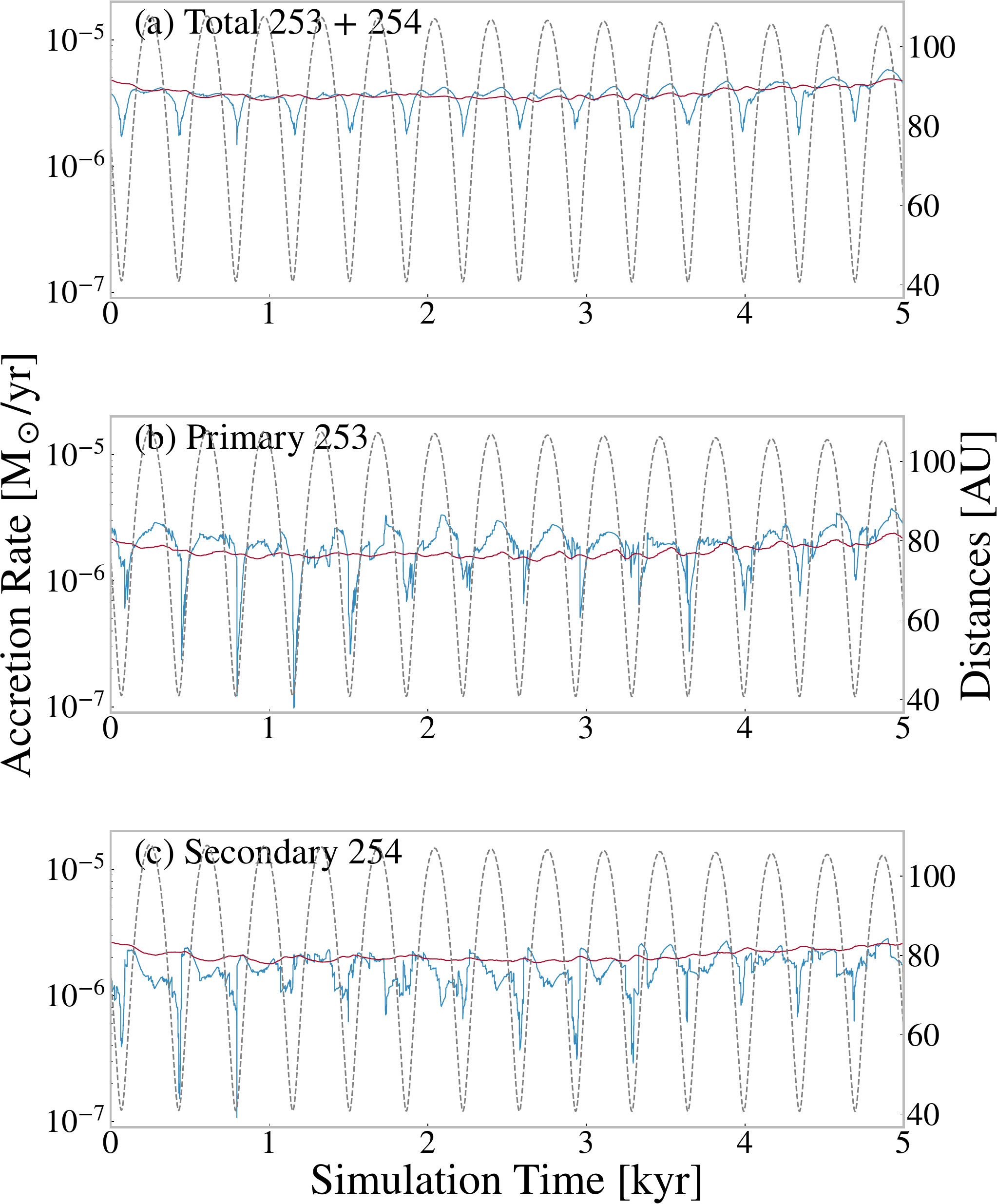}
    \hspace{0.01\columnwidth}
    \includegraphics[width=0.3\columnwidth]{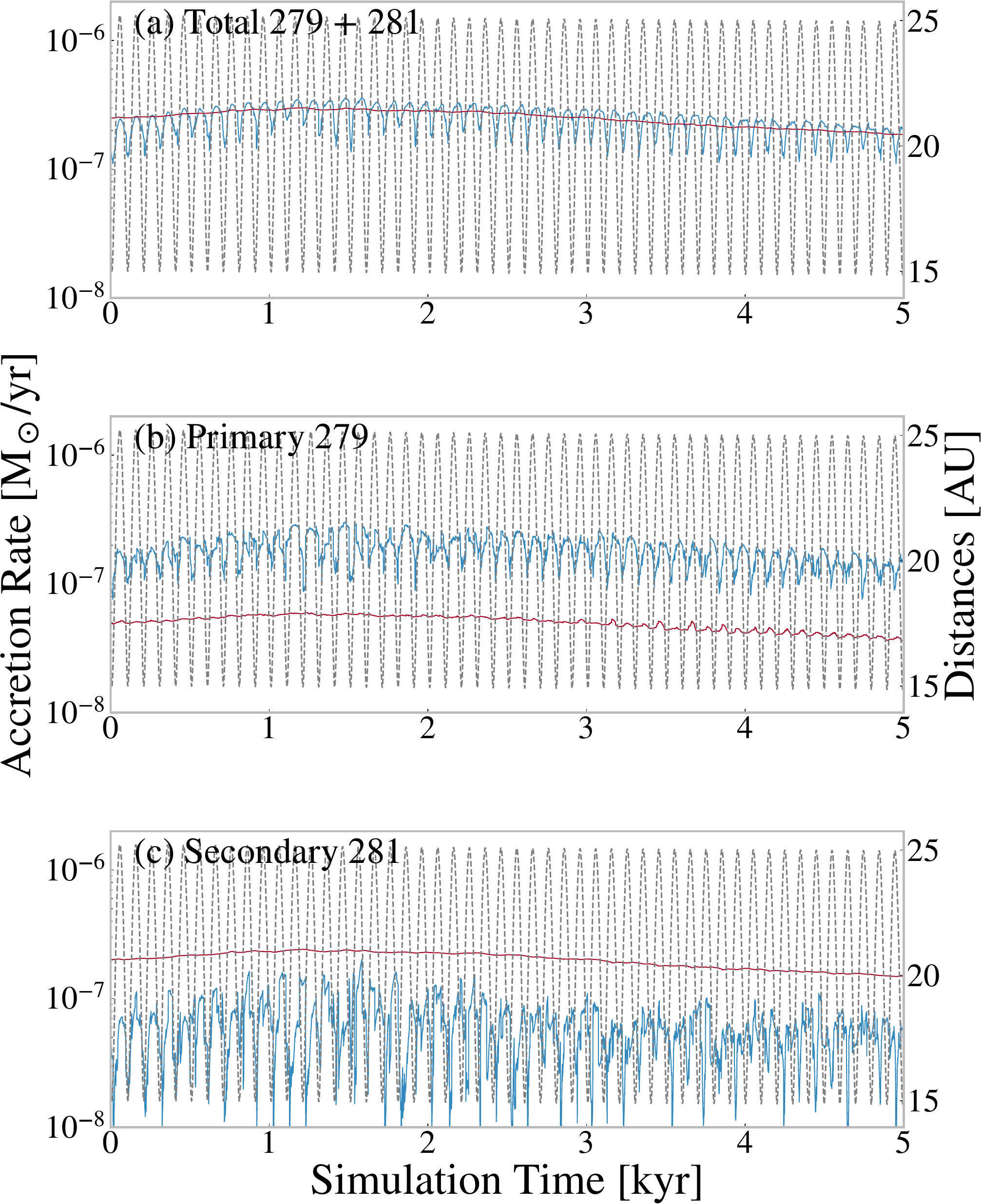}    
    \hspace{0.01\columnwidth}
    \includegraphics[width=0.3\columnwidth]{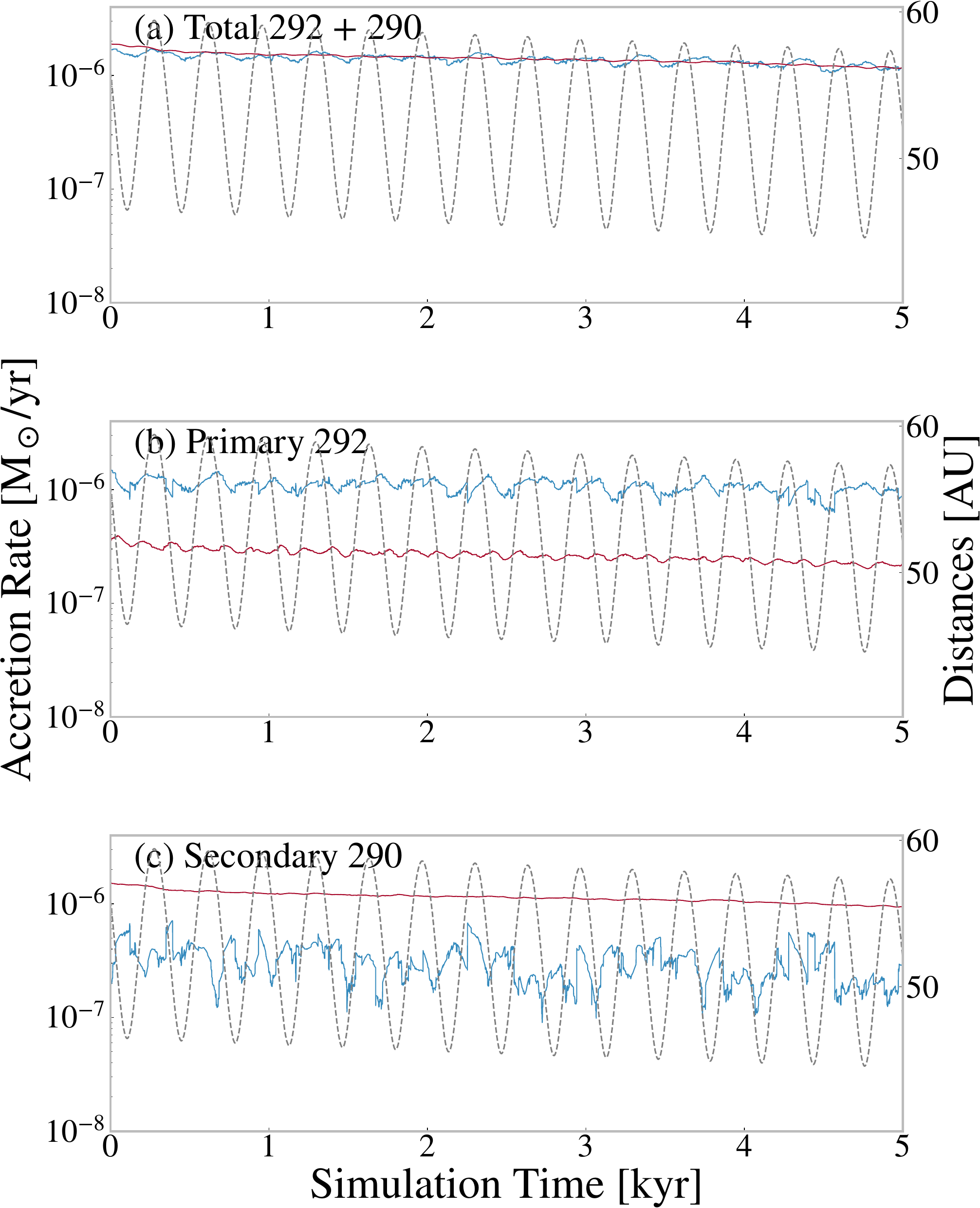}
    
    \captionof{figure}{Same as Fig.~\ref{fig: before-after accretion}, but for sink pairs 75/82, 141/146, 142/144, 217/219, 253/254, 279/281, and 292/290.}
    \label{fig: before-after accretion-A}
\end{center}
\end{minipage}

\clearpage

\begin{minipage}{\textwidth}
\begin{center}
    \hspace{0.635\columnwidth}
    \includegraphics[width=0.3\columnwidth]{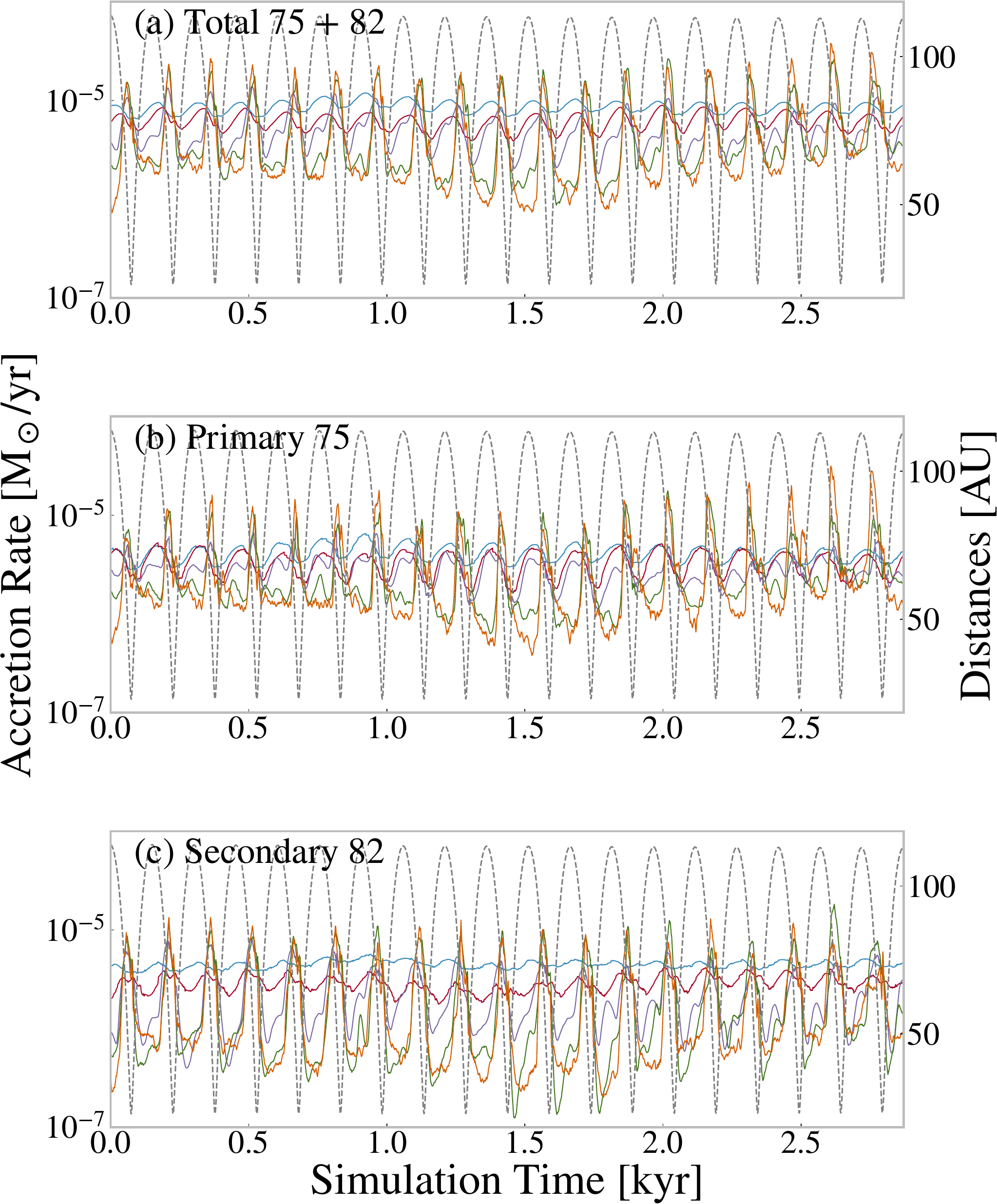}
    \vspace*{0.02\textheight}
    
    \includegraphics[width=0.3\columnwidth]{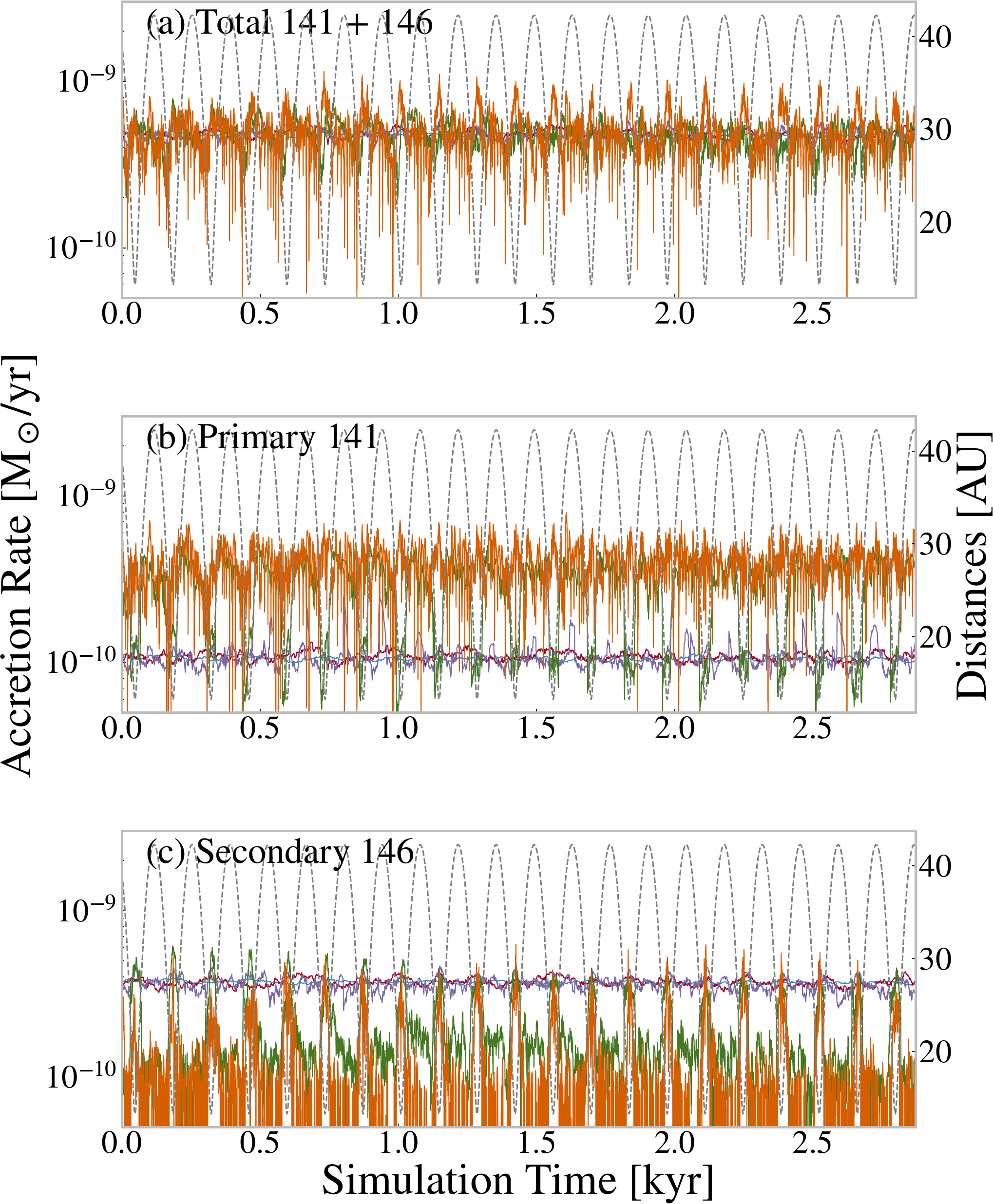}
    \hspace{0.01\columnwidth}
    \includegraphics[width=0.3\columnwidth]{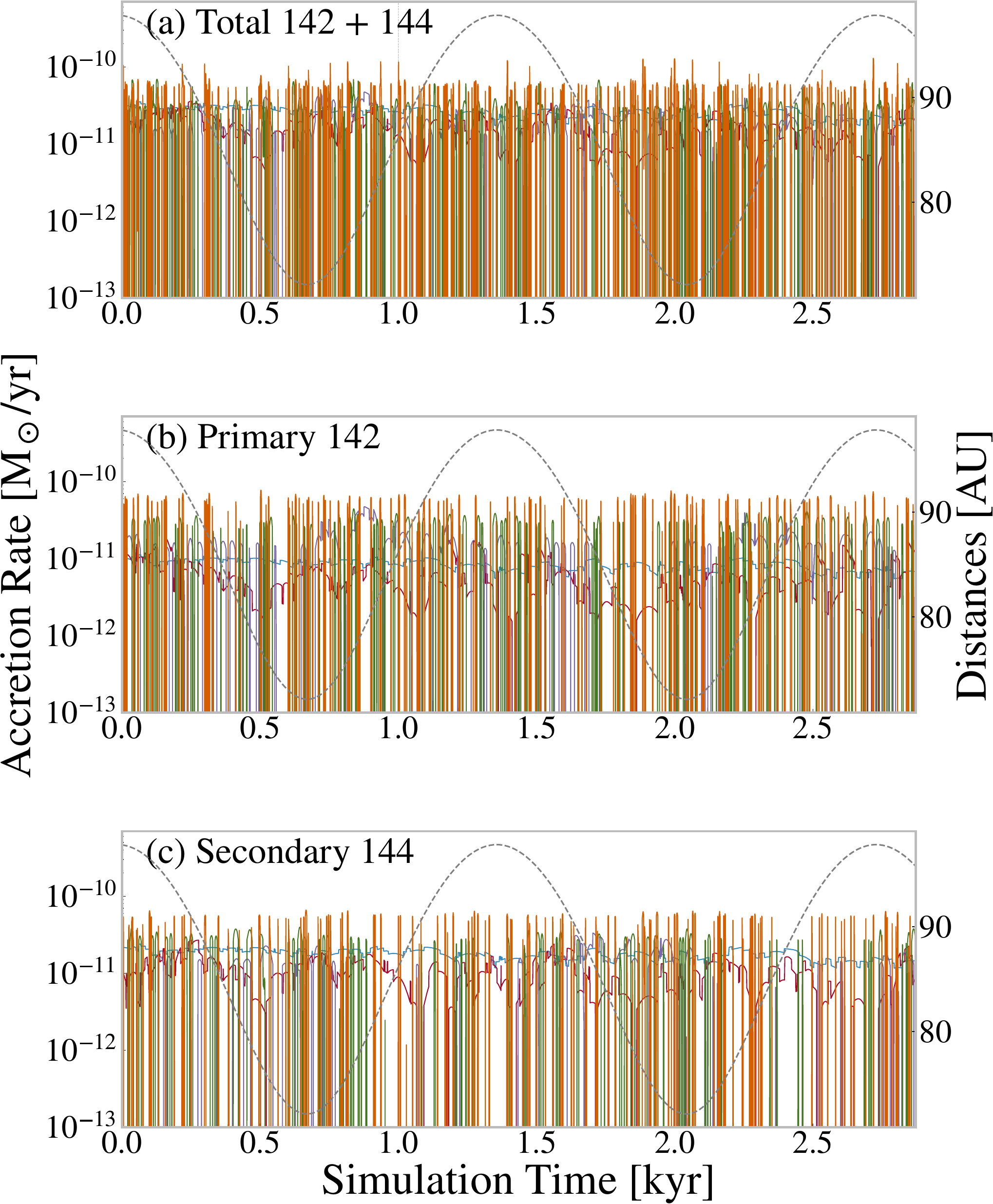}
    \hspace{0.01\columnwidth}
    \includegraphics[width=0.3\columnwidth]{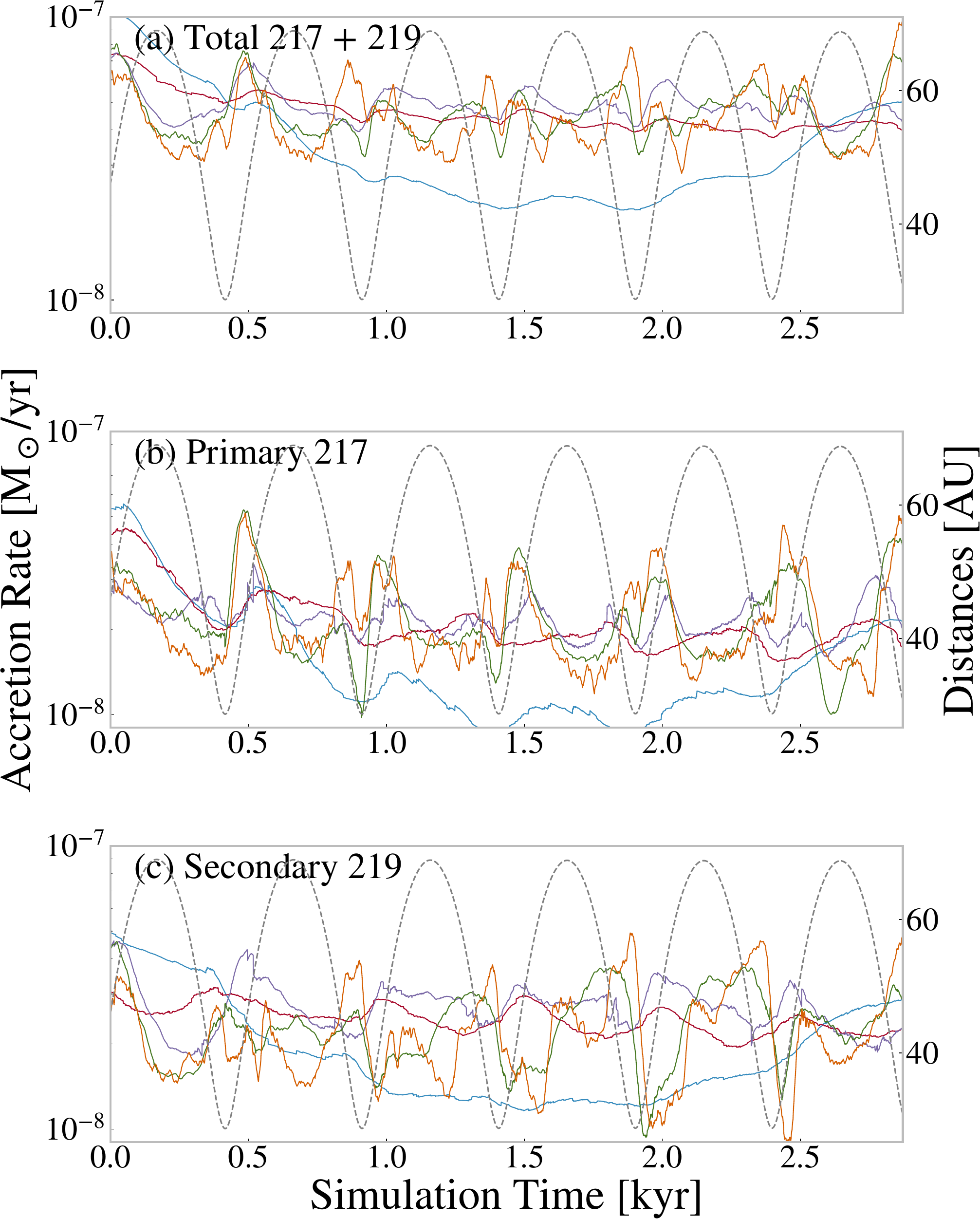}
    \vspace*{0.02\textheight}
    
    \includegraphics[width=0.3\columnwidth]{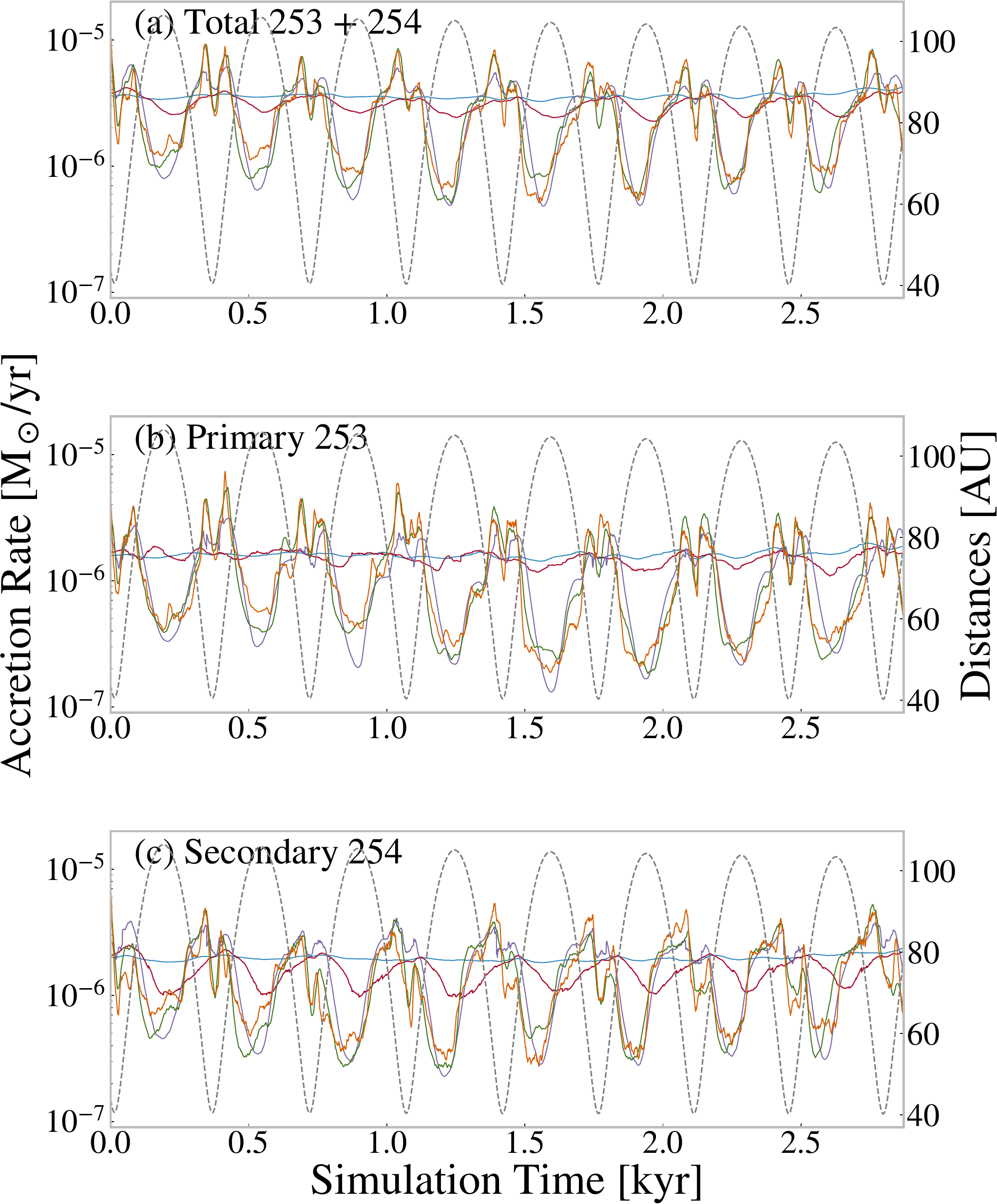}
    \hspace{0.01\columnwidth}
    \includegraphics[width=0.3\columnwidth]{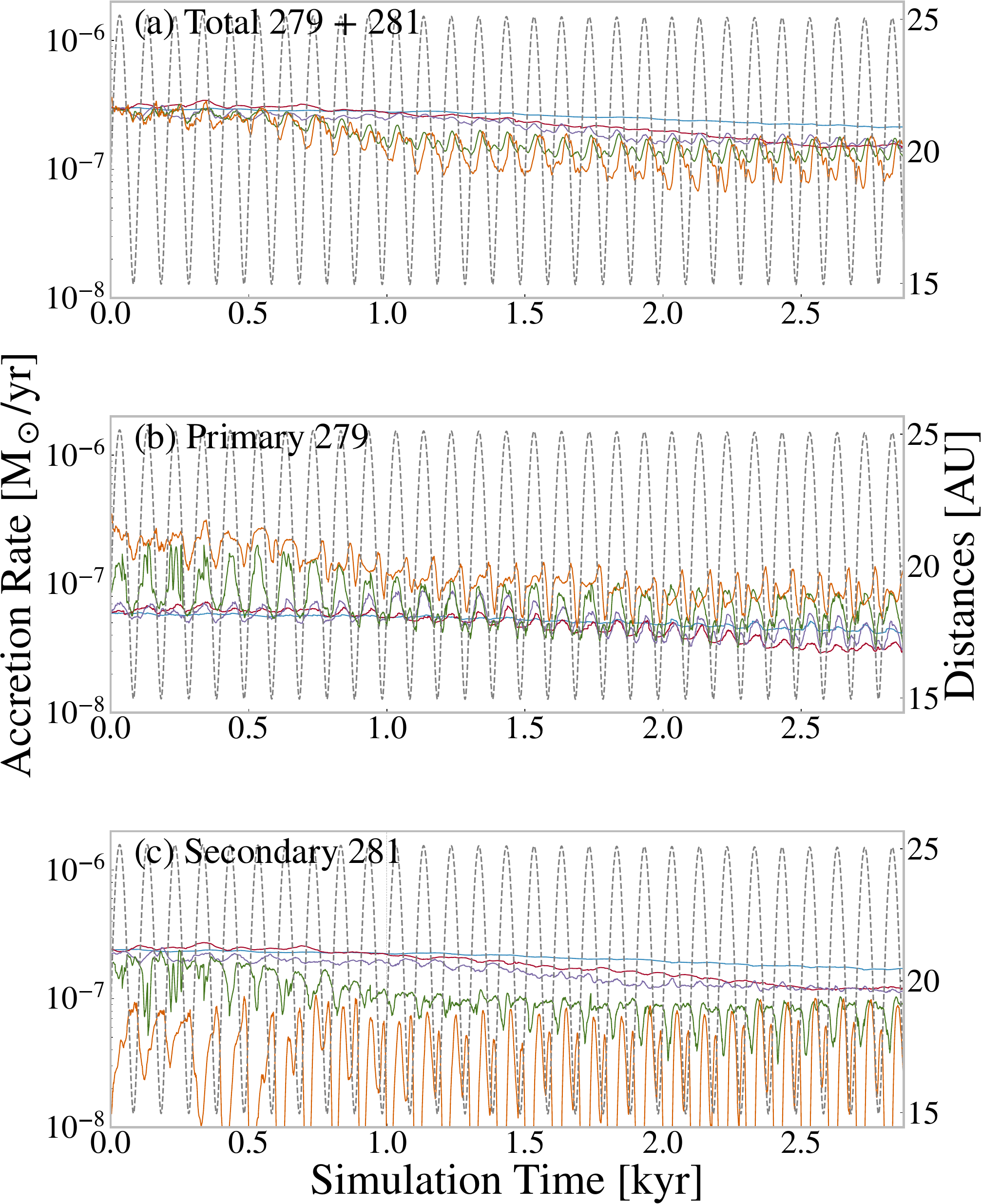}
    \hspace{0.01\columnwidth}
    \includegraphics[width=0.3\columnwidth]{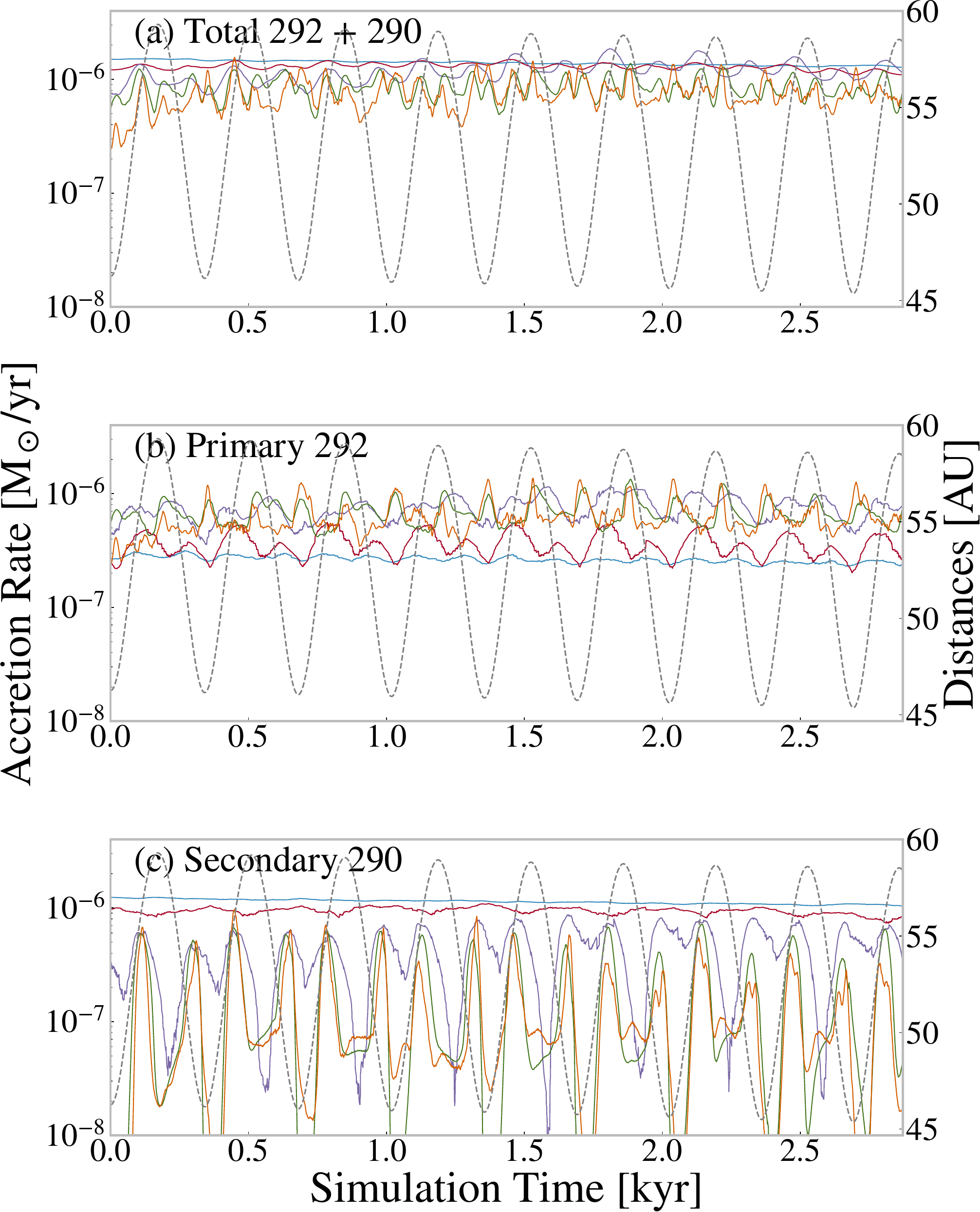}
    
    \captionof{figure}{Same as Fig.~\ref{fig: 257-260 accretion}, but for sink pairs 75/82, 141/146, 142/144, 217/219, 253/254, 279/281, and 292/290.}
    \label{fig: accretion-A}
\end{center}
\end{minipage}

\end{appendix}
\end{document}